\documentclass[journal]{IEEEtran}

\usepackage{amsmath, amssymb, amsthm, graphicx, cite, bbm}

\newtheorem{theorem}{Theorem}
\newtheorem{lemma}{Lemma}
\newtheorem{proposition}{Proposition}
\newtheorem{corollary}{Corollary}
\newtheorem*{result}{Main Result}

\newtheorem{example}{Example}

\newcommand{\Rsum}{R_{\textsf{sum}}}
\newcommand{\Rind}{R_{\textsf{ind}}}

\title{The CEO Problem with $r$th Power of Difference and Logarithmic Distortions}

\author{Daewon Seo and Lav R.~Varshney
\thanks{D.~Seo is with the Department of Electrical Engineering and Computer Science, DGIST, Daegu 42988, South Korea (e-mail: dwseo@dgist.ac.kr). L.~R.~Varshney is with the Coordinated Science Laboratory and the Department of Electrical and Computer Engineering, University of Illinois at Urbana-Champaign, Urbana, IL 61801 USA (e-mail: varshney@illinois.edu).}
\thanks{This work was completed while D.~Seo was at the University of Illinois at Urbana-Champaign. This paper was presented in part at the 2018 IEEE International Symposium on Information Theory (ISIT) \cite{SeoV2019_CEO_isit} and is based in part on a thesis submitted in partial fulfillment of the requirements for the degree of Doctor of Philosophy in Electrical and Computer Engineering at the University of Illinois at Urbana-Champaign.}
\thanks{This work was supported in part by the National Science Foundation under grant CCF-1717530.}
}

\date{}

\begin{document}

\maketitle

\begin{abstract}
	The CEO problem has received much attention since first introduced by Berger \textit{et al}., but there are limited results on non-Gaussian models with non-quadratic distortion measures. In this work, we extend the quadratic Gaussian CEO problem to two non-Gaussian settings with general $r$th power of difference distortion. Assuming an identical observation channel across agents, we study the asymptotics of distortion decay as the number of agents and sum-rate, $\Rsum$, grow without bound, while individual rates vanish. The first setting is a \emph{regular} source-observation model with $r$th power of difference distortion, which subsumes the quadratic Gaussian CEO problem, and we establish that the distortion decays at $\mathcal{O}(\Rsum^{-r/2})$ when $r \ge 2$. We use sample median estimation after the Berger-Tung scheme for achievability. The other setting is a \emph{non-regular} source-observation model, including uniform additive noise models, with $r$th power of difference distortion for which estimation-theoretic regularity conditions do not hold. The distortion decay $\mathcal{O}(\Rsum^{-r})$ when $r \ge 1$ is obtained for the non-regular model by midrange estimator following the Berger-Tung scheme. We also provide converses based on the Shannon lower bound for the regular model and the Chazan-Zakai-Ziv bound for the non-regular model, respectively. Lastly, we provide a sufficient condition for the regular model, under which quadratic and logarithmic distortions are asymptotically equivalent by an entropy power relationship as the number of agents grows. This proof relies on the Bernstein-von Mises theorem.
\end{abstract}


\section{Introduction}
Consider a multiterminal source coding problem where the CEO (Chief Executive Officer) of an organization is interested in a sequence of random variables $\{X(t)\}_{t=1}^{\infty}$, but does not observe it directly. Instead, there are $L$ agents of the organization who make noisy observations; the $i$th agent has $\{Y_i(t)\}_{t=1}^{\infty}$, a noisy version of $\{X(t)\}_{t=1}^{\infty}$ observed over channels with common statistics. The agents must convey their observations to the CEO without convening, but the CEO has a cognitive constraint that limits the information rate she can receive, requiring each agent to discretize observations under rate constraints $\{R_i\}_{i=1}^L$ with $\Rsum = \sum_{i=1}^L R_i$. For a given distortion measure $d$, the CEO declares $\{\hat{X}(t)\}_{t=1}^{\infty}$ that minimizes an expected distortion function $\mathbb{E}[d(X,\hat{X})]$ in a long-term average sense.

The CEO problem was first proposed by Berger \textit{et al.}~\cite{BergerZV1996} for discrete alphabets, and the minimal probability of error was studied when the number of agents and sum-rate tend to infinity. Later a jointly Gaussian setting with quadratic distortion was studied \cite{ViswanathanB1997}, where the asymptotic (in the number of agents) tradeoff between sum-rate and distortion was obtained using a statistical inference approach. The asymptotic result was refined by Oohama \cite{Oohama1998} from the multiterminal source coding viewpoint. The exact rate region of the non-asymptotic quadratic Gaussian CEO problem with a finite number of agents that allows non-identical observation channels was obtained separately by Oohama \cite{Oohama2005} and Prabhakaran \textit{et al.}~\cite{PrabhakaranTR2004}. Under logarithmic distortion, the exact rate region for a general setting with a finite number of agents was found \cite{CourtadeW2014}, and the rate region for the quadratic scalar Gaussian case was explicitly given using quadratic-logarithmic distortion duality \cite{SeoV2016}. The rate region of the vector Gaussian CEO problem under logarithmic distortion is fully characterized \cite{UgurAZ2020}. Several extensions of the quadratic Gaussian CEO problem have been further studied \cite{ChenZBW2004, ChenB2008, WangCW2010, YangX2012, TavildarV2005, EkremU2014}.

In contrast to the quadratic Gaussian CEO problem, non-quadratic and/or non-Gaussian CEO problems have received less attention due to limited analytic tractability compared to the quadratic Gaussian case. A non-regular source-observation pair such as additive uniform noise or truncated Gaussian noise was considered under quadratic distortion \cite{VempatyV2015b}, and a general continuous source with additive Gaussian noise was considered under quadratic distortion and general distortion \cite{EswaranG2019}. Toward a generalization of network compression problems, it was shown that Gaussianity is in fact the worst (that is, least compressible) among variance-bounded sources and variance-bounded additive noises \cite{AsnaniSAW2015}. Beyond CEO problems, a multiterminal setting of hypothesis testing under communication constraints is also widely investigated, but such investigations in continuous-alphabet settings assume that some or all distributions in the problem are Gaussian \cite{TianC2008, RahmanW2012, ZaidiA2019, Zaidi2020}. Specifically, \cite{Zaidi2020} considers independence testing under a general non-Gaussian source, but assumes additive Gaussian noise.

It is common to study quadratic distortion $|x-\hat{x}|^2$ due to its analytic tractability in many cases. For example, there are many useful results such as the Fisher information and the (Bayesian) Cram\'{e}r-Rao lower bound \cite{VanTrees1968}, its mutual information representation in Gaussian channels \cite{GuoSV2005}, linear estimation optimality for Gaussian cases, and so on. However, one might wish to consider lower- or higher-order difference distortion, that is, $|x-\hat{x}|^r$ with $r = 1$ or $r \ge 3$. This is a common extension to the quadratic distortion in the lossy source coding literature e.g., \cite{YamadaTG1980, Algazi1966, BucklewW1982, LinderZ1994, NaN1995}.

In this work, we explore two CEO problems that differ from prior results in that the models have not only a non-Gaussian source-observation pair, but also general $r$th power difference distortion $d(x,\hat{x}) = |x-\hat{x}|^r$. There are two classical asymptotic approaches that have been developed for the CEO and multiterminal source coding problems. The first takes asymptotics in the number of agents \cite{BergerZV1996, ViswanathanB1997}, where the number of agents grows, but individual rates vanish sufficiently slowly so that the sum-rate grows without bound. In this asymptotic regime, the nature of detection or estimation (for discrete and continuous alphabets, respectively) dominates and statistical inference techniques play a key role. The other takes asymptotics where individual coding rate vanishes with a fixed number of agents \cite{ZamirB1999}, which highlights the nature of compression. Note that distortion asymptotics of the two regimes in terms of sum-rate could be different even for the same source-observation setting. In this work, we take the first approach and investigate problems from an estimation-theoretic viewpoint. The models and our contributions are summarized here.

\textbf{Sec.~\ref{sec:regular_ceo}, \emph{regular} model:} Continuous source supported on $\mathbb{R}$ and observation satisfying some regularity conditions, including the jointly Gaussian CEO problem \cite{ViswanathanB1997, Oohama1998, Oohama2005, PrabhakaranTR2004, UgurAZ2020, TavildarV2005, EkremU2014}, but with $d(x, \hat{x})=|x-\hat{x}|^r, r \in \mathbb{N}, r \ge 2$: We find the distortion scales as $\mathcal{O}(\Rsum^{-r/2})$ using the Berger-Tung scheme \cite{Tung1978} with median estimation. A converse bound is also provided by the Shannon lower bound \cite{YamadaTG1980}, assuming each agent makes use of an identical encoding scheme. The tightness of the converse is unknown in general as it reduces to an open problem in information geometry, but it is tight at least when the model is Gaussian.

\textbf{Sec.~\ref{sec:nonregular_ceo}, \emph{non-regular} model:} Bounded source and observation such that estimation-theoretic regularity conditions do not hold, including additive uniform noise, with $d(x, \hat{x})=|x-\hat{x}|^r, r \in \mathbb{N}, r \ge 1$: We find the distortion scales as $\mathcal{O}(\Rsum^{-r})$ using the Berger-Tung scheme with midrange estimation \cite{Rider1957}. A converse bound is also given by a generalization of the Chazan-Zakai-Ziv bound \cite{ChazanZZ1975, Bell1995}, assuming each agent makes use of an identical encoding scheme, although its tightness is unproven.

\textbf{Sec.~\ref{sec:equivalence}, asymptotic equivalence:} If the optimal compression scheme for the regular model in Sec.~\ref{sec:regular_ceo} satisfies some conditions, quadratic (i.e., $r=2$) and logarithmic distortions are asymptotically equivalent as $L \to \infty$, bridged by entropy power equation $D_{\textsf{Q}} = \tfrac{1}{2\pi e} e^{2D_{\textsf{Log}}}$, where $D_{\textsf{Q}}, D_{\textsf{Log}}$ indicate quadratic and logarithmic distortions, respectively. If so, it also implies the logarithmic distortion decays as $\mathcal{O}(-\log \Rsum)$.

This paper is organized as follows. Sec.~\ref{sec:model} formally defines the CEO problem and the figure of merit we are interested in. Sec.~\ref{sec:regular_ceo} states the regular CEO problem and results with complete proofs. Sec.~\ref{sec:nonregular_ceo} states the non-regular CEO problem and results; proof steps that are identical to their regular counterparts are omitted. Sec.~\ref{sec:equivalence} shows the asymptotic equivalence of quadratic and logarithmic distortions. Sec.~\ref{sec:discussion} concludes the paper and mentions a few possible extensions.

\section{CEO Problem Formulation} \label{sec:model}
\begin{figure}[t]
\centering
\includegraphics[width=.45\textwidth]{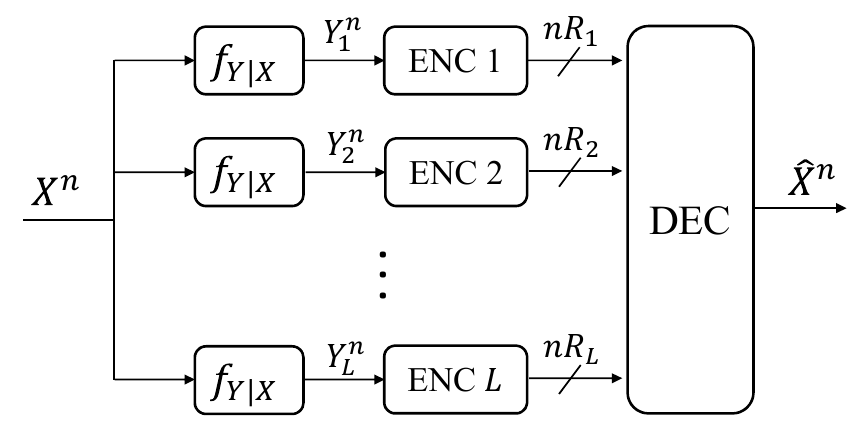}
\caption{The CEO problem model with $L$ agents.}
\label{fig:ceo_topology}
\end{figure}

We consider the CEO problem with real-valued alphabet as in \cite{ViswanathanB1997}, i.e., $\mathcal{X},\mathcal{Y} \subset \mathbb{R}$. The source $\{X(t)\}_{t=1}^{\infty}$ is independent and identically distributed (i.i.d.) from a density function $f_X(x)$. There are $L$ agents who collect the source information, but the $i$th agent is only given a noisy observation $\{Y_i(t)\}_{t=1}^{\infty}$, i.i.d.~drawn from a common observation distribution $f_{Y|X}$. The agents encode observations separately into messages of rate $\{R_i\}_{i=1}^{L}$; more precisely, the $i$th agent encodes length $n$ observations into a codeword $C_i$ from codebook $\mathcal{C}_i$ of rate $R_i$ and sends the codeword index. Sum-rate of the link to the CEO is limited to $\Rsum = \sum_{i=1}^L R_i$.

Upon receiving codewords from agents, the CEO makes estimates $\{\hat{X}(t)\}_{t=1}^{n}$ that minimize the expected distortion of length $n$,
\begin{align*}
D^n(X^n, \hat{X}^n) &:= \frac{1}{n} \mathbb{E} \left[ \sum_{t=1}^{n} |X(t)-\hat{X}(t)|^r \right] \\
&= \frac{1}{n} \sum_{t=1}^{n} \mathbb{E} \left[ |X(t)-\hat{X}(t)|^r \right]
\end{align*}
where $\hat{X} \in \mathcal{X}$, if the distortion is difference distortion of order $r \in \mathbb{N}$. The other distortion measure in this work is logarithmic distortion, which commonly arises in machine learning and also recently in information theory \cite{CourtadeW2014},
\begin{align*}
D_{\textsf{Log}}^n(X^n, \hat{X}^n) &:= \frac{1}{n} \mathbb{E} \left[ \sum_{t=1}^{n} - \log \hat{X}(X; t) \right] \\
&= \frac{1}{n} \sum_{t=1}^{n} \mathbb{E} \left[ - \log \hat{X}(X; t) \right],
\end{align*}
where $\hat{X}$ is a probability distribution over $\mathcal{X}$, i.e., $\hat{X} \in \mathcal{P}(\mathcal{X})$ where $\mathcal{P}(\mathcal{X})$ is the set of absolutely continuous probability distributions (with respect to the Lesbegue measure) over $\mathcal{X}$ that have well-defined differential entropy. The problem model is illustrated in Fig.~\ref{fig:ceo_topology}.

In this work, we are interested in the asymptotic tradeoff between $\Rsum$ and $D^n(X^n,\hat{X}^n)$ with the number of agents in the sense of \cite{BergerZV1996, ViswanathanB1997}. To this end, define 
\begin{align*}
D^n(L,\Rsum) &:= \min_{\{\mathcal{C}_i\}_{i=1}^L: \sum_{i=1}^L R_i \le \Rsum} D^n(X^n, \hat{X}^n), \\
D(L,\Rsum) &:= \lim_{n \to \infty} D^n(L,\Rsum).
\end{align*}
It is immediate that $D(L,\Rsum)$ asymptotically vanishes as $\Rsum, L$ grow without bound. In particular, we consider the regime where individual rates vanish slower than $1/L$ as in \cite{BergerZV1996, ViswanathanB1997}, that is, $\Rsum \to \infty$ and
\begin{align*}
\Rind := \frac{\Rsum}{L} \to 0 \textrm{ as } L \to \infty.
\end{align*}
We investigate the following quantities in this work:
\begin{align*}
\beta_{\textsf{reg}} &:= \lim_{L, \Rsum \to \infty} \Rsum^{r/2} D(L, \Rsum) ~~~ \textrm{in Sec.~\ref{sec:regular_ceo}}, \\
\beta_{\textsf{n-reg}} &:= \lim_{L, \Rsum \to \infty} \Rsum^{r} D(L, \Rsum) ~~~ \textrm{in Sec.~\ref{sec:nonregular_ceo}}.
\end{align*}
If $\beta_{\textsf{reg}}$ and $\beta_{\textsf{n-reg}}$ are positive, the speeds of distortion decay are clearly $\Rsum^{-r/2}$ and $\Rsum^{-r}$ up to some prefactors for regular and non-regular models, respectively.

Before proceeding to formal definitions of regular and non-regular models, recall one of the regularity conditions of the Fisher information (and thus the Cram\'{e}r-Rao lower bound) \cite[Sec.~2.5]{LehmannC2006}:

\begin{itemize}
\item[] The support of $f_{Y|X}$ is common for all $x$, i.e., the set $\{y: f_{Y|X}(y|x) > 0\}$ is independent of $x$.
\end{itemize}
In this context, a model is called \emph{regular} if it satisfies the above condition as well as others in Sec.~\ref{sec:regular_ceo}, whereas called \emph{non-regular} if the above does not hold, but conditions in Sec.~\ref{sec:nonregular_ceo} hold. Note that these two definitions do not form a disjoint partition of the entire model space; some examples are neither regular nor non-regular.

\begin{table*}[t]
	\caption{Glossary} \label{tab:glossary}
	\centering
	\begin{tabular}{ l || l }
		\hline
		symbol & meaning \\ \hline \hline
		$X$, $\hat{X}$ & source signal to be recovered, estimate of $X$ \\ \hline
		$Y_i$ & noisy observation at the $i$th agent \\ \hline
		$C_i, \mathcal{C}_i$ & codeword sent by the $i$th agent, codebook of the $i$th agent \\ \hline
		$r$ & power of difference in distortion \\ \hline
		$D, D_{\textsf{Q}}, D_{\textsf{Log}}$ & $r$th power of difference, quadratic, logarithmic distortions \\ \hline
		$R_i, \Rsum$ & individual coding rate, sum-rate \\ \hline
		$L$ & the number of agents \\ \hline
		$I_Y(x)$ & Fisher information of $Y$ about $x$ \\ \hline
		$\tilde{V}, q(V)$ & discretized value of $V$ \\ \hline
		$\textsf{med}(V)$ & population median of $f_{V}$ \\ \hline
		$\textsf{med}(\{V_i\}), V_{(m+1)}$ & sample median of $\{V_1, \ldots, V_{2m+1}\}$ \\ \hline
		$\mathcal{S}_{\textsf{reg}}$ & set of `good' test channels for the regular model \\ \hline
		$\mathcal{S}_{\textsf{n-reg}}$ & set of `good' test channels for the non-regular model \\
		\hline
	\end{tabular}
\end{table*}

\textbf{Notation:} In the sequel, $f, p$ denote continuous and discrete probability distributions, respectively. We will use the natural logarithm so that the unit of all information quantities are nats. Hat notation $\hat{Z}$ is for estimated values and tilde notation $\tilde{Z}$ is for quantized values. The function $q(\cdot)$ also stands for the quantization function so $q(Z)$ and $\tilde{Z}$ are interchangeable. The round bracket subscript $Z_{(i)}$ denotes the $i$th order statistics, that is, the $i$th element of the reordered sequence from $\{Z_i\}_{i=1}^L$ in increasing order $Z_{(1)} \le Z_{(2)} \le \cdots \le Z_{(L)}$. When $L$ is odd and $L=2m+1$, $Z_{(m+1)}$ is the sample median and denoted by $\textsf{med}(\{Z_i\}_{i=1}^L)$. Also, the population median of $f_Z$ is denoted by $\textsf{med}(Z)$ with abuse of notation. For notational brevity, $Z^n := (Z(1), \ldots, Z(n))$ denotes a length $n$ vector and $Z_{-t}^n := (Z(1), \ldots, Z(t-1), Z(t+1), \ldots, Z(n))$ denotes $Z^n$ excluding the $t$th element. Notation is summarized in Table~\ref{tab:glossary}.

\subsection{Distortion with Varying $r$}
If one considers two different power distortions, say the $r$th power and $p$th power of difference distortions, clearly the distortion characterization varies with the power since units of distortion are power-wise different. One may further wonder how exactly their optimal characterization is different. The following elementary analysis suggests the distortion \emph{might} be different at least in its exponent and this exponent linearly increases with $r$. Our work verifies this forthcoming intuition: the exponents of the upper and lower bounds linearly increase with $r$.

To see this, fix a specific coding and estimation scheme, and recall the definition of the $r$th power of difference distortion.
\begin{align*}
	D_r := \frac{1}{n} \sum_{t=1}^n \mathbb{E}\left[ |X(t) - \hat{X}(t)|^r \right].
\end{align*}
Using Jensen's inequality twice, we have an inequality between $D_r, D_p$ for $1 < p < r$.
\begin{align*}
	D_r &:= \frac{1}{n} \sum_{t=1}^n \mathbb{E}\left[ |X(t) - \hat{X}(t)|^r \right] \\
	&= \frac{1}{n} \sum_{t=1}^n \mathbb{E}\left[ \left( |X(t) - \hat{X}(t)|^p \right)^{r/p}  \right] \\
	&\stackrel{(a)}{\ge} \frac{1}{n} \sum_{t=1}^n \left( \mathbb{E}\left[ |X(t) - \hat{X}(t)|^p  \right] \right)^{r/p} \\
	&\stackrel{(b)}{\ge} \left( \frac{1}{n} \sum_{t=1}^n \mathbb{E}\left[ |X(t) - \hat{X}(t)|^p  \right] \right)^{r/p} = D_p^{r/p},
\end{align*}
where (a) and (b) follow from Jensen's inequality since $(\cdot)^{r/p}$ is convex.

The above inequality indicates that if we can achieve optimal $D_r$ by a certain scheme, $D_p$ incurred by the same estimation scheme, which is perhaps suboptimal for $D_p$, is smaller than $D_r^{p/r}$. In other words, a lower power of difference distortion is immediately upper bounded by $D_r$ with linearly-scaled exponent. Then, an immediate question is whether the equality asymptotically holds or not. That is, whether the linearly scaled exponent is (asymptotically in the number of agents) optimal or not. The linear exponent is not tight if there exists a better estimation scheme that achieves better asymptotics. Our result suggests that the exponent linearly increasing with $r$ is asymptotically (in the number of agents) correct and can be achieved by Berger-Tung coding with median (for the regular model) and midrange estimation (for the non-regular models) schemes.

\section{Regular CEO Problem} \label{sec:regular_ceo}
\subsection{Model and Result}
Consider unbounded source and observation alphabets $\mathcal{X} = \mathcal{Y} = \mathbb{R}$ and impose some regularity conditions on probability distributions that enable us to characterize bounds on $\beta_{\textsf{reg}}$. Let us first state conditions (A1)--(A4) on the source and observation, as well as condition (A5) on the coding scheme. Detailed explanation and intuition for the conditions are deferred until after the main theorem statement, but note that many {\em smooth} distributions satisfy the conditions.
 
\begin{itemize}
\item[(A1)] The source and observations have finite absolute moment of order $r$.

\item[(A2)] The source is identifiable from observations, i.e., $f_{Y|x_1} = f_{Y|x_2}$ only when $x_1 = x_2$.

\item[(A3)] The density $f_X$ is continuous and positive almost everywhere in $\mathbb{R}$. The density $f_{Y|X}$ is twice continuously differentiable with respect to $x$ for almost every $x, y \in \mathbb{R}$. Moreover, for almost every $x \in \mathbb{R}$, the Fisher information
\begin{align*}
I_Y(x):=\mathbb{E}_{Y|x}[(\tfrac{\partial}{\partial x} \log f_{Y|X}(Y|x))^2]
\end{align*}
is well-defined and positive finite for almost every $x \in \mathbb{R}$.
\end{itemize}

Recall that $\textsf{med}(Y|x)$ is the median of $f_{Y|X=x}$, i.e., 
\begin{align*}
\int_{-\infty}^{\textsf{med}(Y|x)} f_{Y|X}(y|x) dy = \frac{1}{2}.
\end{align*}

\begin{itemize}
\item[(A4)] The inverse of $\textsf{med}(Y|x)$ is well-defined and Lipschitz continuous.\footnote{The Lipschitness throughout this paper is on $\mathbb{R}$ equipped with the Euclidean distance.}
\end{itemize}

We impose identical compression schemes on agents in the converse argument.
\begin{itemize}
\item[(A5)] Compression schemes adopted by agents are identical, that is, given the same $y^n$, all agents output the same codewords.
\end{itemize}
Therefore, our converse under condition (A5) gives a conditional lower bound that is proven to hold only when optimal codebooks $\mathcal{C}_i$ are identical across agents. As far as we know, it is unknown whether identical encoding minimizes distortion for CEO problems with $\{Y_i^n\}$ drawn through a common channel, however, identical encoding is optimal for the quadratic Gaussian CEO problem \cite{ChenZBW2004} as well as the discrete CEO problem \cite{BergerZV1996}.

In this work, we use the Berger-Tung coding scheme for achievability, which involves finding an appropriate auxiliary random variable $U$. Define $\mathcal{S}_\textsf{reg}$ to be the set of `good' $U$'s (therefore, good test channels) that satisfy:
\begin{itemize}
	\item[1)] The Markov chain $X-Y-U$ holds.
	
	\item[2)] $U$ has a finite absolute moments of order up to $2r$.
	
	\item[3)] Let $\ell(x):=\textsf{med}(U|x)$ be the function that maps $x$ to the median of $f_{U|x}$. Then, the inverse function $\ell^{-1}(\cdot)$ exists and is Lipschitz with constant $K_U$.
\end{itemize}
Note that $U=Y+V$ with an independent Gaussian $V$ satisfies the three conditions, and therefore, $\mathcal{S}_\textsf{reg}$ is nonempty.

As mentioned, the distortion measure we will consider is the $r$th power of difference, i.e.,
\begin{align*}
d(x,\hat{x}) = |x-\hat{x}|^r, r \in \mathbb{N}, r \ge 2
\end{align*}
under which our main result of this section is the following. Let $\alpha_U := \inf_{x \in \mathbb{R}} f_{U|x}(\textsf{med}(U|x)| x)$.
\begin{result}[Regular CEO problem]
	Suppose conditions (A1)--(A5) hold. Then, for $d(x,\hat{x}) = |x-\hat{x}|^r$, $r \ge 2$, the following bounds hold.
	\begin{align*}
	&C_1 \left( \lim_{a \to 0} \min_{ \substack{U:\\ X-Y-U \\ I(Y;U|X)=a }} \frac{I(Y;U|X)}{ e^{\mathbb{E}[\log I_U(X)]} } \right)^{r/2} \\
	&\le \beta_{\textsf{reg}} \le C_2 \left( \lim_{a \to 0} \min_{ \substack{U \in \mathcal{S}_{\textsf{reg}}:\\ I(Y;U|X)=a }} \frac{K_U^2 I(Y;U|X)}{\alpha_U^2} \right)^{r/2},
	\end{align*}
	where 
	\begin{align*}
	C_1 = \frac{1}{re} \left( \frac{\sqrt{\pi e}}{\sqrt{2} \Gamma\left(1+\frac{1}{r}\right)} \right)^{r}, ~~~ C_2 = \left( \frac{1}{\sqrt{2}}\right)^r \frac{\Gamma(\frac{r+1}{2})}{\sqrt{\pi}}
	\end{align*}
	with the Gamma function $\Gamma(\cdot)$.
\end{result}
\begin{IEEEproof}
	Follows from Thms.~\ref{thm:regular_ceo_ach} and \ref{thm:regular_converse}.
\end{IEEEproof}

\subsection{Discussion on Conditions}
Consider (A1)--(A3) first. Condition (A1) is standard in rate-distortion formulations \cite{Berger1971}; otherwise any arbitrary small error probability in encoding and decoding results in unbounded distortion. Equivalently, (A1) requires that tails of $f_X, f_Y$ decay sufficiently fast. The Cauchy distribution is a well-known example that does not have bounded moments, and therefore, it violates (A1). Condition (A2) is immediate since the CEO is unable to distinguish $x_1$ from $x_2$ unless (A2) holds. Condition (A3) is a regularity condition for Fisher information \cite{LehmannC2006}, so we call this model regular. If (A3) holds, the model is sufficiently {\em smooth} so that we have a closed-form expression for $I(X;Y^L)$ when $L$ is large as in Lem.~\ref{lem:barron_lem}. This is a key idea in our converse argument.

Condition (A4) describes a graphical relationship between $x$ and $\textsf{med}(Y|x)$ that allows us to control estimation error. For illustration, let us consider an example that also gives an overview of our achievability scheme. Suppose $Y_i = X + N_i$, $i=1, \ldots, L$ for large $L$ with i.i.d.~zero mean and unskewed noise $N_i$. Suppose the CEO has direct access to $\{Y_i\}$ and wishes to estimate the source by the sample median. Given $X$, note that $\textsf{med}(\{Y_i\}) \approx X$ with high probability. Therefore, it is reasonable to declare $\hat{X} = \textsf{med}(\{Y_i\})$. Another interpretation of this example is by $\ell(x) := \textsf{med}(Y|x) = x$. As $\ell^{-1}(y) = y$, it is reasonable to declare $\hat{X} = \ell^{-1}(\textsf{med}(\{Y_i\})) = \textsf{med}(\{Y_i\})$. Now consider another example where $Y_i = 2X + N_i$. Similarly, taking $\hat{X} = \ell^{-1}(\textsf{med}(\{Y_i\})) = \tfrac{1}{2} \textsf{med}(\{Y_i\})$ is reasonable because $\ell(x) = 2x$ and $\ell^{-1}(y) = \tfrac{1}{2} y$. Note that $\ell^{-1}$ is Lipschitz with constant $\tfrac{1}{2}$, i.e.,
\begin{align*}
|\ell^{-1}(y+\epsilon) - \ell^{-1}(y) | \le \frac{1}{2} |(y+\epsilon) - y| = \frac{1}{2} \epsilon \quad \textrm{for } \epsilon > 0.
\end{align*}
So $\epsilon$ deviation of the sample median from the population median implies at most $\tfrac{\epsilon}{2}$ deviation of our estimator from the true source. Hence, we can couple two errors in sample median and source recovery.

Now consider a case where $\ell^{-1}$ is not Lipschitz. Suppose $Y_i=\textsf{sgn}(X)\sqrt{|X|}+N_i$ where $X,N_i$ are standard Gaussian and $\textsf{sgn}(\cdot)$ is the sign function. As the noise is additive and unskewed, $\textsf{med}(Y|x) = \textsf{sgn}(x)\sqrt{|x|}$ and the inverse is $\ell^{-1}(z) = \textsf{sgn}(z) z^2$, which is not Lipschitz. Note that when $L$ is large, the sample median $\textsf{med}(\{Y_i\})$ is close to $\textsf{med}(Y|x)$ with high probability, but still has a small error $\epsilon > 0$, say $\textsf{med}(\{Y_i\}) = \textsf{med}(Y|x) + \epsilon = \textsf{sgn}(x)\sqrt{|x|} + \epsilon$. Then, the final estimation error is	
\begin{align*}
|\hat{X} - x| &= |\ell^{-1}(\textsf{med}(\{Y_i\})) - x| \\
&= |\textsf{sgn}(\textsf{med}(\{Y_i\})) (\textsf{med}(\{Y_i\}))^2 - x| \\
&= |\textsf{sgn}(x) (\textsf{sgn}(x)\sqrt{|x|} + \epsilon)^2 - x| \\
&= 2 \epsilon \sqrt{|x|} + \epsilon^2.
\end{align*}
The error grows without bound as $x$ tends to infinity so we fail to accurately recover when $x$ is large. This argument illustrates why Lipschitz continuity of the inverse function is necessary and will be useful in our achievability argument.

Next, consider conditions on $\mathcal{S}_\textsf{reg}$. The Markov condition is required for the Berger-Tung encoding and decoding that we will use. Finite moments are necessary to diminish estimation error occurred by encoding and decoding error events. For instance, the Berger-Tung scheme induced by an additive Cauchy test channel will never achieve a bounded estimation error. The Lipschitz continuity together with (A4) describes a graphical relationship between $x$ and $\textsf{med}(U|x)$ and enables us to control estimation error when no error in encoding and decoding occurred.

Some readers may notice that an independent and additive noise model greatly simplifies the above conditions. Suppose $Y=X+N$ where $N$ is independent noise such that (A1) holds. Then, (A2) immediately holds as $Y=x+N$. As $\textsf{med}(Y|x) = x+ \textsf{med}(N) = x + \textrm{constant}$, the inverse is Lipschitz with constant $1$. So (A4) is also redundant. Only smoothness condition (A3) must be additionally satisfied.

If one adopts another estimation scheme such as mean estimation or maximum likelihood estimation, different conditions need to be imposed. It is however worth mentioning that not only do (A1)--(A5) hold for the model of the Gaussian CEO problem \cite{ViswanathanB1997}, but also that the additive Gaussian test channel is in $\mathcal{S}_\textsf{reg}$.

\subsection{Direct Coding Theorem} \label{subsec:regular_ach}
We will follow the standard achievability approach as in \cite{ViswanathanB1997} except for the median estimator part. That is, first finely quantize continuous alphabets and apply Berger-Tung encoding and decoding over induced discrete alphabets. Then, estimate the source by sample median estimator. Suppose the number of agents is odd for simple exposition, i.e., $L = 2m+1, m \in \mathbb{Z}_+$. Consider random variables $\{U_i\}_{i=1}^L$ generated from an identical test channel $f_{U|Y} \in \mathcal{S}_{\textsf{reg}}$.

\subsubsection{Quantization} \label{subsubsec:regular_quantization}
Quantizing the real line enables agents to use subsequent codes and the Slepian-Wolf (or SW, for short) compression in a discrete domain. Let $\tilde{X}, \tilde{Y}, \tilde{U}$ denote the quantized versions of $X, Y, U$. Our fine quantization ensures that the loss due to quantization is negligible. Formally, we take a quantization scheme that satisfies the following conditions: for some small $\delta_i > 0, i \in \{0,1,2\}$,
\begin{align}
\mathbb{E}\left[ |U_{(m+1)} - q(U_{(m+1)})|^j \right] &\le \delta_0, \label{eq:quant_assumption0} \\
\mathbb{E}\left[ |q(U_{(m+1)}) - \textsf{med}( \{ \tilde{U}_i \}_{i=1}^L ) |^j \right] &\le \delta_0 ~~~ 1 \le j \le 2r, \label{eq:quant_assumption1} \\
|I(Y;U) - I(\tilde{Y};\tilde{U}) | &\le \delta_1, \label{eq:quant_assumption2} \\
|I(X;U) - I(\tilde{X};\tilde{U}) | &\le \delta_2. \label{eq:quant_assumption3}
\end{align}
It is easy to see that there exists a quantization scheme with finite cardinality that satisfies \eqref{eq:quant_assumption0}--\eqref{eq:quant_assumption3}; \eqref{eq:quant_assumption0} and \eqref{eq:quant_assumption1} from the finite moment condition, \eqref{eq:quant_assumption2} and \eqref{eq:quant_assumption3} from the definition of mutual information for arbitrary ensembles \cite{Dobrushin1959}, hence a common refinement of quantization schemes satisfies all conditions. This quantization also induces discrete probability distributions for $\tilde{X}, \tilde{Y}, \tilde{U}$:
\begin{align*}
p_{\tilde{Y}, \tilde{U}} (\tilde{y}, \tilde{u}) &= \int_{\{(y,u):q(y)=\tilde{y}, q(u)=\tilde{u}\}} f_{Y,U}(y,u)dydu, \\
p_{\tilde{X}, \tilde{U}} (\tilde{x}, \tilde{u}) &= \int_{\{(x,u):q(x)=\tilde{x}, q(u)=\tilde{u}\}} f_{X,U}(x,u)dxdu, \\
p_{\tilde{Y}|X}(\tilde{y}|x) &= \int_{\{y:q(y) = \tilde{y}\}} f_{Y|X}(y|x) dy, \\
p_{\tilde{U}|\tilde{Y}}(\tilde{u}|\tilde{y}) &= \frac{ p_{\tilde{Y}, \tilde{U}}(\tilde{y}, \tilde{u}) }{ p_{\tilde{Y}}(\tilde{y}) }.
\end{align*}
Spaces of $\tilde{X}, \tilde{Y}, \tilde{U}$ are denoted by $\tilde{\mathcal{X}}, \tilde{\mathcal{Y}}, \tilde{\mathcal{U}}$, which are of finite cardinality.

\subsubsection{Codes Approximating Test Channel} \label{subsubsec:regular_test_ch}
Each agent takes block length $n_0$ and encodes quantized observation $\tilde{Y}^{n_0}$ into a codeword, instead of $Y^{n_0}$. Let $\varphi:\tilde{\mathcal{Y}}^{n_0} \mapsto \tilde{\mathcal{U}}^{n_0}$ be a deterministic block code encoder, common for all agents. This mapping induces the following empirical distributions,
\begin{align*}
\hat{p}_{\tilde{Y}^{n_0}, \tilde{U}^{n_0}} (\tilde{y}^{n_0}, \tilde{u}^{n_0}) &= p_{\tilde{Y}^{n_0}} (\tilde{y}^{n_0}) \mathbbm{1}_{\{ \varphi(\tilde{y}^{n_0}) = \tilde{u}^{n_0} \}}, \\
\hat{p}_{\tilde{Y}, \tilde{U}}(\tilde{Y}(t) = \tilde{y}, \tilde{U} = \tilde{u}) &= \mathbb{E}_{p_{\tilde{Y}^n}} \left[ \mathbbm{1}_{\{ \tilde{U}(t) = \tilde{u}, \tilde{Y}(t) = \tilde{y} \}} \right], \\
\hat{p}_{\tilde{U}|\tilde{Y}}(\tilde{U}(t) = \tilde{u} | \tilde{Y}(t) = \tilde{y}) &= \frac{\hat{p}_{\tilde{Y}, \tilde{U}}(\tilde{Y}(t) = \tilde{y}, \tilde{U}(t) = \tilde{u})}{p_{\tilde{Y}}(\tilde{Y}(t) = y)},
\end{align*}
where $\mathbbm{1}_{\{\cdot\}}$ is the indicator function. Then the existence of a block code that approximates the true test channel $f_{U|Y}$ follows from \cite[Prop.~3.1]{ViswanathanB1997}.
\begin{proposition}[\cite{ViswanathanB1997}]
For every $\epsilon, \delta > 0$, there exists a deterministic mapping $\varphi:\tilde{Y}^{n_0} \mapsto \tilde{U}^{n_0}$ with the range cardinality $M$ such that 
\begin{align*}
\frac{1}{n_0} \log M &\le I(Y;U) + \epsilon
\end{align*}
and
\begin{align*}
\sum_{\tilde{u} \in \tilde{\mathcal{U}}} |\hat{p}_{\tilde{U}|X}(\tilde{U}(t) = \tilde{u}|x) - p_{\tilde{U}|X}(\tilde{U}(t) = \tilde{u}|x)| &\le \frac{\epsilon}{|\tilde{\mathcal{X}}|}.
\end{align*}
for all $t \in [1:n_0]$ and all $x \in \mathbb{R}$.
\end{proposition}

\subsubsection{Encoding and Decoding} \label{subsubsec:enc_dec}
The overall encoding scheme is two-step as in \cite{BergerZV1996, ViswanathanB1997}: In the first step, each agent encodes $\tilde{Y}_i^{n_0}$ into $\tilde{U}_i^{n_0}$ by the common $\varphi(\cdot)$. Note that the $\{\tilde{U}_i^{n_0}\}$ are correlated since they stem from the same source $X^{n_0}$. Hence, defining supersymbol $W_i := {U}_i^{n_0}$, we can notice that $\{W_i\}_{i=1}^L$ are correlated across agents. In the second step, collecting the $\{W_i\}$ over $n_1$ times, we have the setting for the Slepian-Wolf coding, by which we can remove the correlation and save the rate as if $\{\tilde{W}_i^{n_1}\}_{i=1}^L = \{(\tilde{U}_i^{n_0})^{n_1}\}_{i=1}^L$ are new correlated random sources of length $n_1$. Formally speaking, the SW encoder at the $i$th agent is a mapping $\xi_i: \mathcal{W}^{n_1} \to \{1, \ldots, N_i\}$. Individual and total rates are therefore defined as
\begin{align*}
R_i &= \frac{1}{n_0 n_1} \log N_i, \\
\Rsum &= \sum_{i=1}^L R_i = \frac{1}{n_0 n_1} \sum_{i=1}^L \log N_i.
\end{align*}
The complete encoder of $i$th agent is given by
\begin{align*}
Z_i := \xi_i \circ \varphi^{n_0}(\tilde{y}_i^{n_0 n_1}) \in \{1,\ldots, N_i\},
\end{align*}
which maps $n= n_0 n_1$ observations into a single codeword.

Note that the second encoding step is an $L$-source SW problem with sources being identically distributed conditoned on $X$. That is, the joint distribution of $(\tilde{U}_{i_1}^{n_0}, \tilde{U}_{i_2}^{n_0}, \ldots, \tilde{U}_{i_k}^{n_0})$ is symmetric. It in trun implies the rate region is symmetric in $R_1, \ldots, R_L$. Furthermore, we can achieve the optimal sum-rate of the SW region using codebooks of rate $R_i = \Rsum/L$ for all $i$.

The CEO performs decoding in reverse: recovers $\{\hat{U}_i^{n_0 n_1} \in \tilde{\mathcal{U}}^{n_0 n_1} \}_{i=1}^L$ from $\{Z_i\}_{i=1}^L$, and then estimates $X$ from $\{\hat{U}_i^{n_0 n_1}\}_{i=1}^L$. 

The next proposition (\cite[Prop.~3.2 and Sec.~III.D]{ViswanathanB1997}) characterizes the average individual rate upper bound in multi- and single-letter mutual information forms, and its probability of error.
\begin{proposition}[\cite{ViswanathanB1997}] \label{prop:SW_decoding}
For every $\epsilon, \lambda > 0$ and $\epsilon' > \epsilon$, there exists sufficiently large $L, n$ and index encoders $\{\xi_i\}_{i=1}^L$ such that 
\begin{align*}
\frac{\Rsum}{L} &\le \frac{1}{n_0} H(\tilde{U}^{n_0} | \tilde{X}^{n_0}) + \epsilon \le I(Y;U|X) + \epsilon', \\
\Pr[ \mathcal{B} ] &\le \lambda,
\end{align*}
where $\mathcal{B}:=\{(\hat{U}_1^{n_0}, \ldots, \hat{U}_L^{n_0}) \ne (\tilde{U}_1^{n_0}, \ldots, \tilde{U}_L^{n_0})\}$ is the error event.
\end{proposition}

\subsubsection{Estimation Upper Bound} \label{subsubsec:regular_estimation}
If the CEO has the exact $\textsf{med}(U|x)$, the population median of $U$ resulting from $x$, she can uniquely determine $x=\ell^{-1}(\textsf{med}(U|x))$. Our goal is therefore to estimate $\textsf{med}(U|x)$ accurately from the decoded $\{\hat{U}_i\}_{i=1}^L$, which in turn implies our estimate of $x$ is accurate according to the Lipschitz continuity.

Before proceeding to the achievability, let us consider two lemmas.
\begin{lemma}[Median Estimator \cite{DavidN2003}] \label{lem:Median_Gaussianity}
Let $F, f$ be the cumulative distribution and density function of $V$. Then, the sample median of $L = 2m+1$ samples follows the density function
\begin{align*}
\Pr[V_{(m+1)} = v] &= \frac{(2m+1)!}{m!m!} (F(v))^m (1-F(v))^m f(v) \\
&= \frac{(F(v))^m (1-F(v))^m}{B(m+1, m+1)} dF(v),
\end{align*}
where $B(\cdot, \cdot)$ is the Beta function, so it is the $\textsf{Beta}(m+1,m+1)$ distribution scaled by $F(v)$. Furthermore, $V_{(m+1)}$ is approximately Gaussian $\mathcal{N}\left(\textsf{med}(V), \frac{1}{4 L f^2(\textsf{med}(V))} \right)$ provided that $L$ is large.
\end{lemma}

\begin{lemma} \label{lem:rth_moment}
Under the notations of Lem.~\ref{lem:Median_Gaussianity}, the following holds: For any $\epsilon > 0$, we can take a large $L$ such that
\begin{align*}
&\mathbb{E}[|V_{(m+1)} - \textsf{med}(V)|^r] \\
&\le \left( \frac{1}{2 L f^2(\textsf{med}(V))} \right)^{r/2} \frac{\Gamma(\frac{r+1}{2})}{\sqrt{\pi}} + \epsilon.
\end{align*}
\end{lemma}
\begin{IEEEproof}
See App.~\ref{app:pf_of_med_gau}.
\end{IEEEproof}

Now we can derive the distortion asymptotics (in number of agents) in terms of $\Rsum$.
\begin{theorem}[Achievability of Regular CEO Problem] \label{thm:regular_ceo_ach}
For $r \ge 2$,
\begin{align*}
\beta_{\textsf{reg}} \le \left( \frac{1}{\sqrt{2}}\right)^r \frac{\Gamma(\frac{r+1}{2})}{\sqrt{\pi}} \left( \lim_{a \to 0} \min_{ \substack{U \in \mathcal{S}_{\textsf{reg}}:\\ I(Y;U|X)=a }} \frac{K_U^2 I(Y;U|X)}{\alpha_U^2} \right)^{\frac{r}{2}}.
\end{align*}
\end{theorem}
\begin{IEEEproof}
Fix $f_{U|Y}$. The expectation of an instantaneous error at $t$ is bounded as follows. As $\ell(x) = \textsf{med}(U|x) \in \mathcal{U}$, we take our estimation $\hat{X} = \ell^{-1}(\hat{U}_{(m+1)}(t))$. Then,
\begin{align*}
&\mathbb{E}\left[ |X(t) - \hat{X}(t)|^r \right] = \mathbb{E} \left[ |X(t) - \ell^{-1}(\hat{U}_{(m+1)}(t))|^r \right] \\
&\stackrel{(a)}{\le} K_U^r \mathbb{E} \left[ |\textsf{med}(U|X) - \hat{U}_{(m+1)}(t)|^r \right] \\
&= K_U^r \mathbb{E} \left[ |\textsf{med}(U|X) - U_{(m+1)}(t) + U_{(m+1)}(t) - \hat{U}_{(m+1)}(t)|^r \right] \\
&\stackrel{(b)}{\le} K_U^r \mathbb{E} \left[ |\textsf{med}(U|X)- U_{(m+1)}(t)|^r \right] + \epsilon_1,
\end{align*}
where (a) follows from the Lipschitz property of $\ell^{-1}$, and (b) follows from the fine quantization and vanishing error probability of the Slepian-Wolf scheme. Details of (b) are relegated to App.~\ref{app:pf_of_epsilon}.

Consider the first term. As the median estimator is approximately $\mathcal{N}\left(\textsf{med}(U|X), \frac{1}{4 L f^2(\textsf{med}(U|X))} \right)$ for large $L$,
\begin{align*}
&K_U^r \mathbb{E} \left[ |\textsf{med}(U|X)- U_{(m+1)}(t)|^r \right] \\
&=K_U^r \mathbb{E}_X \mathbb{E}_{U|X} \left[ |\textsf{med}(U|X)- U_{(m+1)}(t)|^r | X \right] \\
&\le K_U^r \mathbb{E}_X \left[ \left( \frac{1}{ 2L f^2(\textsf{med}(U|X))} \right)^{r/2} \frac{\Gamma(\frac{r+1}{2})}{\sqrt{\pi}} + \epsilon_2 \Big| X \right] \\
&= \left(\frac{K_U}{\sqrt{2} \alpha_U}\right)^r \frac{1}{L^{r/2}} \frac{\Gamma(\frac{r+1}{2})}{\sqrt{\pi}} + \epsilon_2,
\end{align*}
where the inequality follows from Lem.~\ref{lem:rth_moment}.

Summing over all $t \in [1:n]$ and letting $\epsilon = \epsilon_1 + \epsilon_2$,
\begin{align*}
D^n(X^n, \hat{X}^n) &= \frac{1}{n} \sum_{t=1}^n \mathbb{E}\left[ |X(t) - \hat{X}(t)|^r \right] \\
&\le \left(\frac{K_U}{\sqrt{2} \alpha_U}\right)^r \frac{1}{L^{r/2}} \frac{\Gamma(\frac{r+1}{2})}{\sqrt{\pi}} + \epsilon \\
\implies D(L,\Rsum) &\le \left(\frac{K_U}{\sqrt{2} \alpha_U}\right)^r \frac{1}{L^{r/2}} \frac{\Gamma(\frac{r+1}{2})}{\sqrt{\pi}} + \epsilon.
\end{align*}

From Prop.~\ref{prop:SW_decoding}, we have $\tfrac{\Rsum}{L} \le I(Y;U|X)$, and therefore,
\begin{align*}
\beta_{\textsf{reg}} &= \lim_{L, \Rsum \to \infty} \Rsum^{r/2} D(L, \Rsum) \\
&\le \left( \frac{1}{\sqrt{2}}\right)^r \frac{\Gamma(\frac{r+1}{2})}{\sqrt{\pi}} \left( \frac{K_U^2 I(Y;U|X)}{\alpha_U^2} \right)^{r/2}.
\end{align*}
Taking infimum over $f_{U|Y} \in \mathcal{S}_{\textsf{reg}}$ such that $I(Y;U|X)=\textrm{constant}$ and decreasing individual rates to zero complete the proof.
\end{IEEEproof}

\subsection{Converse Coding Theorem} \label{subsec:regular_converse}
A key idea of the converse is the Shannon lower bound \cite{YamadaTG1980}. For any source and distortion measure, the classic rate-distortion function is given in \cite{Shannon1959, Berger1971}; however, a closed-form expression of the rate-distortion function is unavailable except for a few special cases. For real-valued sources and difference-norm distortion, the Shannon lower bound is useful not only because of a simple closed-form expression, but also because of asymptotic tightness when distortion is small \cite{LinderZ1994, Koch2016}. Forthcoming Lem.~\ref{lem:barron_lem} stems from \cite{ClarkeB1990, ClarkeB1994} to characterize mutual information growth when the number of observations tends to infinity.

We consider a genie-aided converse, where the decoder is given $X_{-t}^n$ at the moment of decoding and estimating $X(t)$. For completeness, we begin with coding rate analysis that follows standard arguments as in \cite{ViswanathanB1997}.

\subsubsection{Coding Rate Lower Bound} \label{subsubsec:coding_rate_LB}
Let us first derive a coding rate lower bound.
\begin{align*}
nR_i &= \log |\mathcal{C}_i^n| \\
&\ge I(Y_i^n;C_i|X_n) \\
&= \sum_{t=1}^n I(Y_i(t);C_i|Y_i^{t-1}, X_n) \\
&= \sum_{t=1}^n \left[ h(Y_i(t)|Y_i^{t-1}, X^n) - h(Y_i(t)|C_i,Y_i^{t-1}, X^n) \right] \\
&= \sum_{t=1}^n \left[ h(Y_i(t)|X^n) - h(Y_i(t)|C_i,Y_i^{t-1}, X^n) \right] \\
&\ge \sum_{t=1}^n \left[ h(Y_i(t)|X^n) - h(Y_i(t)|C_i,X^n) \right] \\
&= \sum_{t=1}^n I(Y_i(t);C_i|X^n).
\end{align*}
A sum-rate lower bound is therefore given by
\begin{align*}
\Rsum \ge \frac{1}{n} \sum_{t=1}^n \sum_{i=1}^L I(Y_i(t);C_i|X^n) = \frac{L}{n} \sum_{t=1}^n I(Y(t);C|X^n),
\end{align*}
where the last equality follows from the identical encoder assumption over agents.

Let $U(t, x_{-t}^n)$ denote a codeword random variable whose joint distribution with $X(t)$ and $Y(t)$ is
\begin{align*}
&\Pr[x \le X(t) \le x+dx, y \le Y(t) \le y+dy, u(t, x_{-t}^n)=c] \\
&= f_X(x) f_{Y|X}(y|x) \\
&~~~ ~~~ ~~~ \times \Pr[C = c|Y(t) = y, X(t) = x, X_{-t}^n = x_{-t}^n] dx dy \\
&= f_X(x) f_{Y|X}(y|x) \Pr[C = c|Y(t) = y, X_{-t}^n = x_{-t}^n] dx dy,
\end{align*}
since codeword $C$ depends on $X(t)$ only through $Y(t)$. Hence, the Markov chain $X(t) - Y(t) - U(t, x_{-t}^n)$ holds, and so we have the following lower bound in expectation form.
\begin{align*}
\Rsum \ge \frac{L}{n} \sum_{t=1}^n \mathbb{E}_{X_{-t}^n} [I(Y(t);U(t, X_{-t}^n)|X(t))].
\end{align*}

\subsubsection{Estimation Lower Bound}
An estimate by the CEO is $\hat{X}^n(C_1, C_2, \ldots, C_L)$. Consider two-component lemmas prior to the estimation lower bound.

\begin{lemma}[Shannon lower bound \cite{YamadaTG1980}] \label{lem:Shannon_LB}
Suppose $X, \hat{X}$ are $d$-dimensional source and estimate vectors in $\mathbb{R}^d$ and consider any norm $\|X-\hat{X}\|$. Define the standard rate distortion function
\begin{align*}
R(D) := \inf_{P_{\hat{X}|X}: \mathbb{E}[ \| X-\hat{X} \|^r] \le D} I(X;\hat{X}).
\end{align*}
Then, the Shannon lower bound is given by
\begin{align*}
R(D) &\ge R_{\textsf{SLB}}(D) := h(X) - \frac{d}{r} \log \left( \frac{rD}{d} (V_d \Gamma(1+\frac{d}{r}))^{r/d} e \right) \\
&= h(X) - \frac{1}{r} \log \left( reD (2 \Gamma(1+\frac{1}{r}))^{r} \right) ~~~\textrm{if } d=1,
\end{align*}
where $V_d$ is the volume of $d$-dimensional unit ball such that $\{ x: \| x \| \le 1, x \in \mathbb{R}^d\}$
and $\Gamma(\cdot)$ is the Gamma function.
\end{lemma}

\begin{lemma}[\cite{ClarkeB1990, ClarkeB1994}] \label{lem:barron_lem}
Suppose $X \in \mathbb{R}^d$ and conditions (A2)--(A3) hold. Then,
\begin{align}
&I(X(t);\{C_i\}_{i=1}^L) \nonumber \\
&= \frac{d}{2} \log \frac{L}{2 \pi e} + h(X) + \frac{1}{2} \mathbb{E}[\log \det I_C(X(t))] + o(1) \nonumber \\
&= \frac{1}{2} \log \frac{L}{2 \pi e} + h(X) + \frac{1}{2} \mathbb{E}[\log I_C(X(t))] + o(1) ~~~\textrm{if } d=1. \label{eq:ClarkeBarron_eq}
\end{align}
\end{lemma}

Now we can prove the converse.
\begin{theorem}[Converse of Regular CEO Problem] \label{thm:regular_converse}
Suppose (A5) holds, i.e., each agent adopts the same codebook. Then, for any conditional density $f_{U|Y}$, it holds that for $r \ge 2$,
\begin{align*}
\beta_{\textsf{reg}} &\ge \frac{1}{re} \left( \frac{\sqrt{\pi e}}{\sqrt{2} \Gamma\left(1+\frac{1}{r}\right)} \right)^{r} \\
&~~~ ~~~ ~~~ \times  \left( \lim_{a \to 0} \min_{ \substack{U:\\ X-Y-U \\ I(Y;U|X)=a }} \frac{I(Y;U|X)}{ e^{\mathbb{E}[\log I_U(X)]} } \right)^{r/2}.
\end{align*}
\end{theorem}
\begin{IEEEproof}
Combining Lems.~\ref{lem:Shannon_LB} and \ref{lem:barron_lem} with $d=1$, we have the following chain of inequalities: For any $D \ge 0$,
\begin{align*}
&h(X) - \frac{1}{r} \log \left( reD (2 \Gamma(1+\frac{1}{r}))^{r} \right) \\
&\stackrel{(a)}{\le} \inf_{P_{\hat{X}|X}: \mathbb{E}[ | X(t)-\hat{X}(t) |^r] \le D} I(X(t);\hat{X}(t)) \\
&\stackrel{(b)}{\le} I(X(t);\{C_i\}_{i=1}^L) \\
&\stackrel{(c)}{=} \frac{1}{2} \log \frac{L}{ 2 \pi e} + h(X) + \frac{1}{2} \mathbb{E}[\log I_{ C }(X(t))] + o(1),
\end{align*}
where (a) follows from the Shannon lower bound in Lem.~\ref{lem:Shannon_LB}, (b) follows from the data processing inequality since a Markov chain $X(t) - \{C_i\}_{i=1}^L - \hat{X}(t)$ holds, and (c) follows from Lem.~\ref{lem:barron_lem}.

Rearranging terms,
\begin{align*}
D(t) &\ge \left(\frac{1}{L e^{\mathbb{E}[\log I_C(X(t))]} }\right)^{r/2} \frac{1}{re} \left( \frac{\sqrt{\pi e}}{\sqrt{2} \Gamma\left(1+\frac{1}{r} \right)} \right)^r \\
&= C_1  \left(\frac{1}{L e^{\mathbb{E}[\log I_C(X(t))]} }\right)^{r/2},
\end{align*}
where
\begin{align*}
C_1 := \frac{1}{re} \left( \frac{\sqrt{\pi e}}{\sqrt{2} \Gamma\left(1+\frac{1}{r}\right)} \right)^{r}.
\end{align*}
Hence,
\begin{align*}
D^n(L,\Rsum) &= \frac{1}{n} \sum_{t=1}^n \mathbb{E}[|X(t) - \hat{X}(t)|^r] \\
&\ge \frac{C_1}{n} \sum_{t=1}^n \left(\frac{1}{L e^{\mathbb{E}[\log I_C(X(t))]} }\right)^{r/2} \\
&= \frac{C_1}{L^{r/2}} \frac{1}{n} \sum_{t=1}^n \left(\frac{1}{e^{\mathbb{E}[\log I_C(X(t))]} }\right)^{r/2} \\
&\stackrel{(d)}{\ge} \frac{C_1}{L^{r/2}} \left( \frac{1}{n} \sum_{t=1}^n \frac{1}{e^{\mathbb{E}[\log I_C(X(t))]} }\right)^{r/2} \\
&\stackrel{(e)}{\ge} \frac{C_1}{L^{r/2}} \left( \frac{n}{\sum_{t=1}^n e^{\mathbb{E}[\log I_C(X(t))]} } \right)^{r/2},
\end{align*}
where (d) follows from Jensen's inequality, and (e) follows from the harmonic mean-arithmetic mean inequality.

Multiplying $D^n(L, \Rsum)$ by $\Rsum^{r/2}$,
\begin{align*}
\beta_{\textsf{reg}} &\ge \left( \frac{L}{n} \sum_{t=1}^n \mathbb{E}_{X_{-t}^n}[I(Y(t);U(t, X_{-t}^n) | X(t))] \right)^{r/2} \\
&~~~ ~~~ ~~~ ~~~ \times \frac{C_1}{L^{r/2}} \left( \frac{n}{\sum_{t=1}^n e^{\mathbb{E}[\log I_C(X(t))]} } \right)^{r/2} \\
&= C_1 \left( \frac{\sum_{t=1}^n \mathbb{E}_{X_{-t}^n}[I(Y(t);U(t, X_{-t}^n) | X(t))]}{\sum_{t=1}^n e^{ \mathbb{E}[\log I_C(X(t))]} } \right)^{r/2} \\
&\ge C_1 \left( \min_{t, X_{-t}^n} \frac{I(Y(t);U(t, X_{-t}^n) | X(t))}{ e^{\mathbb{E}[\log I_C(X(t))]} } \right)^{r/2}.
\end{align*}
Therefore we obtain a single-letter bound
\begin{align*}
\beta_{\textsf{reg}} \ge C_1 \left( \min_{U} \frac{I(Y;U | X)}{ e^{\mathbb{E}[\log I_U(X)]} } \right)^{r/2}.
\end{align*}
Decreasing $I(Y;U | X) \to 0$ concludes the converse.
\end{IEEEproof}

Further applying Jensen's inequality yields the generalization of the converse by Viswanathan and Berger \cite{ViswanathanB1997}.
\begin{corollary} \label{cor:VB_LB}
	For any conditional density $f_{U|Y}$, it holds that for $r \ge 2$,
	\begin{align*}
	\beta_{\textsf{reg}} \ge \frac{1}{re} \left( \frac{\sqrt{\pi e}}{\sqrt{2} \Gamma\left(1+\frac{1}{r}\right)} \right)^{r} \left( \lim_{a \to 0} \min_{ \substack{U:\\ X-Y-U \\ I(Y;U|X)=a }} \frac{I(Y;U|X)}{ \mathbb{E}[I_U(X)] } \right)^{\frac{r}{2}}.
	\end{align*}
\end{corollary}
\begin{IEEEproof}
The only new term is the ratio $\frac{I(Y;U|X)}{\mathbb{E}[I_U(X)]}$. Since the exponential function is convex, Jensen's inequality yields
\begin{align*}
e^{\mathbb{E}[\log I_U(X)]} \le \mathbb{E}[e^{\log I_U(X)}] = \mathbb{E}[I_U(X)].
\end{align*}
Therefore the claim holds.
\end{IEEEproof}

Notice that the term $\frac{I(Y;U|X)}{\mathbb{E}[I_U(X)]}$ is the ratio of the conditional mutual information to the Fisher information. The setting maximizing the Fisher information, while having mutual information fixed, is an important problem in information theory \cite{Amari1989, HanA1995}. Our lower bound also has a similar expression that maximizes the Fisher information when the conditional mutual information vanishes.

Investigating whether the lower bound is strictly positive is not straightforward: as far as we know it is still open. However, for the Gaussian CEO model, it reduces to \cite{ViswanathanB1997, Oohama1998} as we will see, thus positive.

\subsection{Examples}

We consider three examples of the regular model.

\begin{example}
	Let us consider the quadratic Gaussian CEO problem \cite{ViswanathanB1997, Oohama1998}. Suppose $X \sim \mathcal{N}(0, \sigma_X^2)$, $Y = X+N$ where $N \sim \mathcal{N}(0, \sigma_N^2)$. Then, Viswanathan-Berger \cite{ViswanathanB1997} and Oohama \cite{Oohama1998} showed for quadratic distortion, i.e., $r=2$, 
	\begin{align*}
		\beta_{\textsf{VB}}:=\lim_{\Rsum, L \to \infty} \Rsum D(L, \Rsum) = \frac{\sigma_N^2}{2}.
	\end{align*}
	Moreover, the limit can be attained by an additive Gaussian compression kernel with unbounded noise variance. That is, $U=Y+V$, $V \sim \mathcal{N}(0, \sigma_V^2)$ and $\sigma_V^2 \to \infty$. Then, sample mean estimation achieves $\beta_{\textsf{VB}}=\sigma_N^2/2$. Evaluating the non-asymptotic (in the number of agents) result \cite{PrabhakaranTR2004, Oohama2005} for our asymptotic (in the number of agents) setting, this conclusion can be directly recovered too.
	
	Returning to our result, note that $C_1 = 1, C_2 = 1/4$ when $r=2$. Then, Thms.~\ref{thm:regular_ceo_ach} and \ref{thm:regular_converse} reduce to
	\begin{align*}
		&\lim_{I(Y;U|X) \to 0} \min_{U} \frac{I(Y;U | X)}{ e^{\mathbb{E}[\log I_U(X)]} } \le \beta_{\textsf{reg}} \\
		&~~~ ~~~ ~~~ ~~~ ~~~ \le \frac{1}{4} \lim_{I(Y;U|X) \to 0} \min_{ U \in \mathcal{S}_{\textsf{reg}}   } \frac{K_U^2 I(Y;U|X)}{\alpha_U^2}.
	\end{align*}
	Minimizers of the lower and upper bounds are hard to obtain in general, however, it is reasonable to take the additive Gaussian compression kernel as \cite{ViswanathanB1997}. From Cor.~\ref{cor:VB_LB}, the lower bound recovers $\beta_{\textsf{VB}}$ since
	\begin{align*}
		\lim_{\sigma_V^2 \to \infty} \frac{I(Y;U|X)}{\mathbb{E}[I_U(X)]} = \lim_{\sigma_V^2 \to \infty} \frac{\frac{1}{2} \log (1+\frac{\sigma_N^2}{\sigma_V^2})}{1/(\sigma_N^2 + \sigma_V^2)} = \frac{\sigma_N^2}{2} = \beta_{\textsf{VB}}.
	\end{align*}
	
	On the other hand, note that $\alpha_U = (2\pi(\sigma_N^2+\sigma_V^2))^{-1/2}$ and $K_U = 1$ due to Gaussianity and additivity. Consequently, the upper bound yields
	\begin{align*}
		\lim_{\sigma_V^2 \to \infty} \frac{1}{4} \frac{I(Y;U|X)}{\alpha_U^2} = \lim_{\sigma_V^2 \to \infty} \frac{1}{4} \frac{\frac{1}{2} \log (1+\frac{\sigma_N^2}{\sigma_V^2})}{(2\pi(\sigma_N^2+\sigma_V^2))^{-1}} = \frac{\pi \sigma_N^2}{4}.
	\end{align*}
	The loss in prefactor $0.2854 \approx \tfrac{\pi}{4}-\tfrac{1}{2}$ is because of the suboptimality of median estimation for Gaussian distributions, however, it provides the correct asymptotics (in number of agents) $\Theta(\Rsum^{-r/2})$. Median estimator often outperforms sample mean estimator in particular when the noise variance is large, but is also robust against adversarial corruption \cite{Huber1981}.
\end{example}

Next two examples evaluate upper bounds for non-Gaussian regular models.
\begin{example}
	Consider a quadratic sub-Gaussian CEO problem $Y=X+N$ where $X, N$ are sub-Gaussian independently drawn from $f(t) = \frac{1}{Z} \exp(-ct^4)$, where
	\begin{align*}
		c = \frac{\Gamma(\frac{3}{4})^4}{2 \pi^2 \sigma^4} \quad \textrm{and} \quad Z = \frac{\pi^{3/2} \sigma}{2^{1/4} \Gamma(\frac{3}{4})^2}
	\end{align*}
	so that $f$ is a valid probability density function with mean zero and variance $\sigma^2$. Suppose $X, N$ have variances $1$ and $\sigma_N^2$ respectively.
	
	For achievability, it is challenging to minimize the right side, but the additive Gaussian test channel is a valid choice for tractability. So take $U = Y+V = X+N+V$ and $V \sim \mathcal{N}(0,\sigma_V^2)$. As Gaussian maximizes differential entropy given variance,
	\begin{align*}
	I(Y;U|X) &= \int f_X(x) I(Y;U|X=x) dx \\
	&= \int f_X(x) (h(N+V) - h(V)) dx \\
	&\le \int f_X(x) \left( \frac{1}{2} \log \left( 1+\frac{\sigma_N^2}{\sigma_V^2} \right) \right) dx \\
	&= \frac{1}{2} \log \left( 1+\frac{\sigma_N^2}{\sigma_V^2} \right).
	\end{align*}
	Moreover, since the noise and test channel both are independent and additive, $K_U=1$ and $\alpha_U = f_{U|X}(0|0) = f_{N+V}(0)$. The convolution property of the sum of two independent random variables gives
	\begin{align*}
		\alpha_U &= f_{N+V}(0) = \int_{-\infty}^{\infty} f_N(-\tau) f_V(\tau) d\tau \\
		&= \frac{1}{Z \sqrt{2\pi \sigma_V^2}} \int_{-\infty}^{\infty} \exp(-c\tau^4) \exp(-\frac{\tau^2}{2\sigma_V^2}) d\tau \\
		&= \frac{e^{1/32\sigma_V^4 c} B_K(\tfrac{1}{4}, \tfrac{1}{32\sigma_V^4 c})}{4 Z \sqrt{\pi c} \sigma_V^2}
	\end{align*}
	where $B_K$ is the modified Bessel function of the second kind. Therefore, we can calculate the limit
	\begin{align*}
		 \lim_{\sigma_V^2 \to \infty} \frac{1}{4} \frac{I(Y;U|X)}{\alpha_U^2} \le \lim_{\sigma_V^2 \to \infty} \frac{1}{4} \frac{\frac{1}{2} \log \left( 1+\frac{\sigma_N^2}{\sigma_V^2} \right)}{\alpha_U^2} =\frac{\pi \sigma_N^2}{4}.
	\end{align*}
	Therefore, $\mathcal{O}(\Rsum^{-1})$ is achievable.
	
	It should be noted that the bound is the same as that we obtained for the quadratic Gaussian case. This is not surprising because we can always upper bound $h(N+V)$ by the Gaussian entropy although $N+V$ is non-Gaussian. Moreover, the test channel is highly noisy Gaussian, the Gaussian nature dominates the behavior of $U$. Therefore, we can achieve at least the same performance of the Gaussian CEO problem for any non-Gaussian regular model. This generalizes the fact that Gaussianity is the worst in compression and communication, i.e., least compressible in source coding or least favorable in communication \cite{AsnaniSAW2015}, to an indirect rate-distortion setting.
\end{example}

\begin{example}
	Consider the standard Gaussian source and additive chi-square noise of degree of freedom $1$, i.e., $Y=X+N$ where $X \sim \mathcal{N}(0,1)$ and $N \sim \chi^2(1)$. We take the additive chi-square test channel $V \sim \chi^2(k)$ due to tractability, and therefore, $U=X+N+V$. As sum of chi-square random variables is still chi-square, we know $N+V \sim \chi^2(k+1)$.
	
	Noting that this is an additive model, $I(Y;U|X=x)$ is invariant over $x$. Hence, the mutual information can be obtained in closed form as
	\begin{align*}
		&I(Y;U|X) = I(Y;U|X=x) = h(N+V) - h(V) \\
		&= \left[ \frac{k+1}{2} + \log\left( 2\Gamma\left( \frac{k+1}{2} \right)\right) + \left( 1-\frac{k+1}{2} \right) \psi\left( \frac{k+1}{2} \right) \right] \\
		&~~~ ~~~ ~~~ - \left[ \frac{k}{2} + \log\left( 2\Gamma\left( \frac{k}{2} \right)\right) + \left( 1-\frac{k}{2} \right) \psi\left( \frac{k}{2} \right) \right],
	\end{align*}
	where $\Gamma(\cdot), \psi(\cdot)$ are the Gamma and digamma functions, respectively.

	In addition, as the population median of the chi-square distribution of degree $k+1$ is approximately $(k+1)(1-\frac{2}{9(k+1)})^3 =: \zeta$, we have
	\begin{align*}
		\alpha_U &= f_{N+V}(\textsf{med}(N+V)) = f_{N+V}\left( \zeta \right) \\
		&= \frac{1}{2^{\frac{k+1}{2}} \Gamma \left(\frac{k+1}{2}\right)} \zeta^{\frac{k+1}{2}-1} \exp \left( -\frac{\zeta}{2}\right).
	\end{align*}

	With the fact that $C_2 = \tfrac{1}{4}, K_U = 1$, we can evaluate the upper bound.
	\begin{align*}
	\beta \le \lim_{k \to \infty} \frac{1}{4} \frac{I(Y;U|X)}{\alpha_U^2} \approx 1.5708.
	\end{align*}
	Hence, $\mathcal{O}(\Rsum^{-1})$ is achievable.
\end{example}

\section{Non-regular Model} \label{sec:nonregular_ceo}
\subsection{Model and Result}
This section considers bounded source and observation alphabets $\mathcal{X}, \mathcal{Y} \subsetneq \mathbb{R}$ as in \cite{VempatyV2015b}, where the source-observation model is assumed to be \emph{non-regular} in the sense of the Cram\'{e}r-Rao lower bound regularity condition \cite{VanTrees1968, LehmannC2006}. An example of such non-regular models is $Y=X+N$ with $X, N \sim \textsf{unif}[0,1]$ independently. Let us first state formal definitions. Detailed explanation and intuition are deferred until after the main theorem statement.
\begin{itemize}
	\item[(B1)] The source and observation are finitely supported, that is, $\mathcal{X}, \mathcal{Y} \subset \mathbb{R}$ are (possibly finite union of) bounded intervals of the real line.
	
	\item[(B2)] $f_{Y|x}$ is discontinuous at both end points of support conditioned on $x$, i.e., let $\mathcal{Y}_x \subset [e_{\ell}(x), e_u(x)]$ be the support of $f_{Y|x}$ for some $-\infty < e_{\ell}, e_u < \infty$, and $\lim_{y \downarrow e_{\ell}(x)} f_{Y|x}(y), ~\lim_{y \uparrow e_u(x)}f_{Y|x}(y) > 0$. Furthermore, $(e_{\ell} + e_u)(x)$ is invertible and the inverse is Lipschitz continuous.
\end{itemize}
Without loss of generality, we suppose $\mathcal{X} = \mathcal{Y} = [0,1]$. 

As the regular model, we impose an identical coding scheme on agents in the converse.
\begin{itemize}
	\item[(B3)] Compression schemes adopted by agents are identical, that is, given the same $y^n$, all agents output the same codeword.
\end{itemize}

Let $\mathcal{S}_{\textsf{n-reg}}$ be the set of good test channels for the non-regular model (i.e., the set of $U$s in the Berger-Tung achievability) that satisfy: 
\begin{itemize}
	\item[1)] A Markov chain $X-Y-U$ holds.
	
	\item[2)] $f_{U|x}$ has a bounded support $\mathcal{U}_x \in [a(x), b(x)]$, and $f_{U|x}$ does not vanish at either end point $a(x), b(x)$, i.e., $f_{U|X}(a(x)|x), f_{U|X}(b(x)|x) > \delta_U$ for some $\delta_U>0$.
	
	\item[3)] $\ell(x):=\tfrac{a+b}{2}(x)$ is invertible and the inverse function $\ell^{-1} = (\tfrac{a+b}{2})^{-1}$ is Lipschitz with constant $K_U$.
\end{itemize}
Without loss of generality, we assume $\mathcal{U} = [0,1]$ as well.

To see that $\mathcal{S}_{\textsf{n-reg}}$ is nonempty, consider a $k \ge 2$ peak noise $V$ such that $f_V(v) = \sum_{i=1}^k p_i \delta(v-c_i)$ where $\delta(\cdot)$ is the Dirac-delta function and $c_1 < c_2 < \cdots < c_k$ are peak locations. If a test channel is taken as $U = Y+V$, the conditional distribution $f_{U|x}(u) = \sum_{i=1}^k p_i f_{Y|x}(u-c_i)$ is supported on $\mathcal{U}_x \subset \cup_i [e_{\ell}(x) + c_i, e_u(x)+c_i]$. In addition,
\begin{align*}
f_{U|x}(e_{\ell} + c_1) &= \lim_{u \downarrow (e_{\ell} + c_1)} f_{U|x}(u) = p_1 \lim_{y \downarrow e_{\ell}(x)} f_{Y|x}(y) > 0, \\
f_{U|x}(e_{u} + c_k) &= \lim_{u \uparrow (e_{u} + c_k} f_{U|x}(u) = p_k \lim_{y \uparrow e_{u}(x)} f_{Y|x}(y) > 0.
\end{align*}
Hence it satisfies all the conditions, $\mathcal{S}_{\textsf{n-reg}}$ is clearly nonempty. Such a discrete test channel is investigated in \cite{Fix1978}.

\begin{result}[Non-regular CEO problem]
	Suppose conditions (B1)--(B3) hold. Then, for $d(x, \hat{x}) = |x-\hat{x}|^r$,$r\ge 1$, the following bounds hold.
	\begin{align*}
	&\frac{r}{2^r} \int_{\tilde{h}=0}^1 \tilde{h}^{r-1} \int_{x=0}^{1} f_X(x) \varphi(x) dx d\tilde{h} \\
	&\le \beta_{\textsf{n-reg}} \le 2(r!) \left( \lim_{a \to 0} \min_{\substack{U \in \mathcal{S}_{\textsf{n-reg}}: \\ I(Y;U|X) =a }} \frac{K_U I(Y;U|X)}{\delta_U} \right)^r,
	\end{align*}
	with
	\begin{align*}
	\varphi(x) &= \lim_{a \to 0} \min_{ \substack{U:\\ X-Y-U \\ I(Y;U|X)=a }} I(Y;U|X)^r e^{-\tilde{h} g(x)}, \\
	g(x) &= \frac{d}{d\Delta} \left( - \min_{s \in [0,1]} \log \left( \int f_{U|X}^s (u|x) f_{U|X}^{1-s} (u|x + \Delta) dy \right) \right) \Bigg\vert_{\Delta=0}.
	\end{align*}
\end{result}

Before proceeding, it should be noted that proofs in the sequel repeat standard parts of achievability and converse proofs in Sec.~\ref{subsec:regular_ach} and Sec.~\ref{subsec:regular_converse} so are omitted.

\subsection{Discussion on Conditions}
It should be noted that the Cram\'{e}r-Rao lower bound \cite[Sec.~2.5]{LehmannC2006} requires that the support of $f_{Y|x}$ be the same for all $x \in \mathcal{X}$, which our model does not satisfy. So we call the model non-regular. Also note that $f_X$ (or $f_{Y|x}$) is allowed to be zero inbetween two intervals on which $f_X$ (or $f_{Y|x}$) is positive.

As illustration, let us consider two simple examples of the non-regular model.
\begin{example} \label{ex4}
	Let $Y = X + N$ where $X, N$ are independently drawn from $\textsf{unif}[0,1]$. Then, $f_{Y|x}$ is supported on $[x, x+1] =: [e_{\ell}(x), e_u(x)]$. So $(e_{\ell}+e_u)(x) = 2x+1$ is invertible and the inverse is Lipschitz. Moreover, $\lim_{y \downarrow x} f_{Y|x}(y) = \lim_{y \uparrow x+1}f_{Y|x}(y) = 1$, non-vanishing.
\end{example}

\begin{example}
	Let $X \sim \textsf{unif}[0,1]$ with observation $Y \sim \textsf{unif}[0,X]$. The support of $f_{Y|x}$ is $[0,x] =:[e_{\ell}(x), e_u(x)]$. Thus, $(e_{\ell}+e_u)(x) = x$ is invertible and the inverse is Lipschitz. Moreover, $\lim_{y \downarrow 0} f_{Y|x}(y) = \lim_{y \uparrow x}f_{Y|x}(y) =\tfrac{1}{x} \ge 1$ for all $x$, non-vanishing.
\end{example}

To illustrate the role of Lipschitz continuity, let us revisit Ex.~\ref{ex4} and suppose the CEO has direct access to $\{Y_i\}$, that is, the additive test channel is $V = 0$ or equivalently $V \sim f_V(v) = \delta(v)$. Hence, the conditional support of $U$ is $\mathcal{U}_x = [x, 1+x]=:[a(x), b(x)]$, and thus, $\ell(x) = \tfrac{a+b}{2}(x) = x+\tfrac{1}{2}$ is invertible with Lipschitz constant $K_U=1$. As $f_{Y|x}(a(x))$ and $f_{Y|x}(b(x))$ are nonzero, there will be some observations in small neighborhoods of $a(x), b(x)$ if $L$ is large. So if the number of observations given to the CEO is large enough, she can accurately estimate $\tfrac{a+b}{2}(x)$ by $\tfrac{\min_i \{Y_i\}+\max_i \{Y_i\}}{2}$. This in turn implies her estimate $\hat{x} = \ell^{-1} \left( \tfrac{\min_i \{Y_i\}+\max_i \{Y_i\}}{2} \right)$ is accurate as well due to the Lipschitz continuity of $\ell^{-1}$.

Recall that $\mathcal{S}_{\textsf{n-reg}}$ includes a collection of discrete test channels. However, $\mathcal{S}_{\textsf{n-reg}}$ does not admit an independent and additive test channel whose density $f_{U|Y}$ is continuous as follows.
\begin{proposition} \label{prop:collection_delta}
	Consider a non-regular model that satisfies (B1) and (B2). Then, an independent and additive test channel such that $f_{U|Y}$ is continuous in $y$ cannot satisfy $f_{U|X}(a(x)|x), f_{U|X}(b(x)|x) >0$. 
\end{proposition}
\begin{proof}
	Given in App.~\ref{app:pf_collection_delta}.
\end{proof}

\subsection{Direct Coding Theorem}
Like in Sec.~\ref{subsec:regular_ach}, we use quantization, Berger-Tung compression-decompression, and then estimating the source. Conditions for the quantization are the following:
\begin{align}
\mathbb{E}[|U-\tilde{U}|^j] &\le \delta_0 ~~~ 1 \le j \le 2r, \label{eq:non_reg_quan} \\
|I(Y;U) - I(\tilde{Y};\tilde{U})| &\le \delta_1, \nonumber \\
|I(X;U) - I(\tilde{X};\tilde{U})| &\le \delta_2. \nonumber
\end{align}
The remaining steps are the same as Sec.~\ref{subsec:regular_ach} except for the estimation step. The midrange estimator will be used to estimate the source since it is optimal in several cases with bounded supports \cite{NeymanP1928, Rider1957, ArceF1988}. Furthermore, it is more efficient than the sample mean in many cases such as the cosine, parabolic, rectangular, and inverted parabolic distributions \cite{Rider1957}.

\begin{theorem}[Achievability of non-regular CEO problem] \label{thm:non-regular_ach}
	For $r \ge 1$,
	\begin{align*}
	\beta_{\textsf{n-reg}} &\le 2(r!) \left( \lim_{a \to 0} \min_{\substack{U \in \mathcal{S}_{\textsf{n-reg}}: \\ I(Y;U|X) =a }} \frac{K_U I(Y;U|X)}{\delta_U} \right)^r,
	\end{align*}
	where $K_U, \delta_U$ are from the definition of $\mathcal{S}_{\textsf{n-reg}}$.
\end{theorem}
\begin{IEEEproof}
	As mentioned, the CEO estimates $X$ by sample midrange estimator, i.e.,
	\begin{align*}
	\hat{X}(t) = \ell^{-1} \left( \frac{\hat{U}_{(1)}(t) + \hat{U}_{(L)}(t)}{2} \right).
	\end{align*}
	Fix $f_{U|Y}$. Then we have the following distortion upper bound:
	\begin{align*}
	& \mathbb{E}\left[ |X(t)-\hat{X}(t)|^r \right] \\
	&= \mathbb{E} \left[ \left| X(t) - \ell^{-1} \left( \frac{\hat{U}_{(1)}(t) + \hat{U}_{(L)}(t)}{2} \right) \right|^r \right] \\
	&= \mathbb{E} \Bigg[ \Bigg| \ell^{-1} \left( \frac{a(X(t)) + b(X(t))}{2} \right) \\
	&~~~ ~~~ ~~~ ~~~ ~~~ ~~~ - \ell^{-1} \left( \frac{\hat{U}_{(1)}(t) + \hat{U}_{(L)}(t)}{2} \right) \Bigg|^r \Bigg] \\
	&\le K_U^r \mathbb{E} \left[ \left| \frac{a(X(t)) + b(X(t))}{2} - \frac{\hat{U}_{(1)}(t) + \hat{U}_{(L)}(t)}{2} \right|^r \right]
	\end{align*}
	since $\ell^{-1}$ is Lipschitz with constant $K$. For notational simplicity, let us denote $a_X = a(X(t)), b_X = b(X(t))$ and omit `$(t)$'.
	\begin{align*}
	& K_U^r \mathbb{E} \left[ \left| \frac{a_X + b_X}{2} - \frac{\hat{U}_{(1)} + \hat{U}_{(L)}}{2} \right|^r \right] \\
	&= \left(\frac{K_U}{2}\right)^r \mathbb{E} \left[ \left| (a_X + b_X) - (\hat{U}_{(1)} + \hat{U}_{(L)}) \right|^r \right] \\
	&= \left(\frac{K_U}{2}\right)^r \mathbb{E} \Big[ \Big| (a_X + b_X) - (U_{(1)} + U_{(L)}) \\
	&~~~ ~~~ ~~~ ~~~ ~~~ ~~~ ~~~ ~~~ ~~~ + (U_{(1)} + U_{(L)}) - (\hat{U}_{(1)} + \hat{U}_{(L)}) \Big|^r \Big]  \\
	&\stackrel{(a)}{\le} \left(\frac{K_U}{2}\right)^r \mathbb{E} \Big[ \Big| |(a_X + b_X) - (U_{(1)} + U_{(L)})| \\
	&~~~ ~~~ ~~~ ~~~ ~~~ ~~~ ~~~ ~~~ ~~~ + |(U_{(1)} + U_{(L)}) - (\hat{U}_{(1)} + \hat{U}_{(L)})| \Big|^r \Big] \\
	&\stackrel{(b)}{\le} \left(\frac{K_U}{2}\right)^r \mathbb{E} \left[ \left|(a_X + b_X) - (U_{(1)} + U_{(L)}) \right|^r \right] + \epsilon_3,
	\end{align*}
	where (a) follows from the triangle inequality, and (b) is given by Prop.~\ref{prop:non-regular_epsilon} in App.~\ref{app:pf_of_epsilon}.
	
	Recall that $f_{U|X}$ does not vanish at either end point, $a_X$ and $b_X$. Define the set $\mathcal{I}:=\{u_{(1)} > a_X+\epsilon_4 \textrm{ or } u_{(L)} < b_X-\epsilon_4 \}$ so that $\mathcal{I}^c = \{U_{(1)} \le a_X + \epsilon_4 \textrm{ and } U_{(L)} \ge b_X-\epsilon_4 \}$. Note that
	\begin{align*}
	&\mathbb{E} \left[ \left| a_X + b_X - U_{(1)} - U_{(L)} \right|^r \right] \\
	&= \mathbb{E}_X \mathbb{E}_{U|X} \left[ \left| a_X + b_X - U_{(1)} - U_{(L)} \right|^r \Big| X \right].
	\end{align*}
	Then, the conditional expectation is
	\begin{align}
	& \mathbb{E}_{U|X} \left[ \left| a_X + b_X - U_{(1)} - U_{(L)} \right|^r \Big| X \right] \nonumber \\
	&= \int_{\mathcal{I}} \left| a_X + b_X - U_{(1)} - U_{(L)} \right|^r \nonumber \\
	&~~~ ~~~ ~~~ ~~~ ~~~ ~~~ \times f_{U_{(1)}, U_{(L)}|X}(u_{(1)}, u_{(L)}|x) du_{(1)} du_{(L)} \nonumber \\
	&~~~ + \int_{\mathcal{I}^c} \left| a_X + b_X - U_{(1)} - U_{(L)} \right|^r \nonumber \\
	&~~~ ~~~ ~~~ ~~~ ~~~ ~~~ \times f_{U_{(1)}, U_{(L)}|X}(u_{(1)}, u_{(L)}|x) du_{(1)} du_{(L)} \nonumber \\
	&\le const \cdot \Pr[\mathcal{I}|X] \nonumber \\
	&~~~ + \int_{\mathcal{I}^c} \left| a_X + b_X - U_{(1)} - U_{(L)} \right|^r \nonumber \\
	&~~~ ~~~ ~~~ ~~~ ~~~ ~~~ \times f_{U_{(1)}, U_{(L)}|X}(u_{(1)}, u_{(L)}|x) du_{(1)} du_{(L)}. \label{eq:ach:error_bound}
	\end{align}
	Let us separately evaluate each term. First, since $\{U_i\}_{i=1}^L$ are independent when conditioned on $X$,
	\begin{align*}
	&\Pr[\mathcal{I}|X] \\
	&\le \Pr[U_{(1)} > a_X + \epsilon_4 |X] + \Pr[U_{(L)} < b_X - \epsilon_4 |X] \\
	&= \prod_{i=1}^L \Pr[U_{i} > a_X + \epsilon_4 |X] + \prod_{i=1}^L \Pr[U_{i} < b_X - \epsilon_4 |X] \\
	&= \prod_{i=1}^L (1-\Pr[U_{i} \le a_X + \epsilon_4 |X]) \\
	&~~~ ~~~ ~~~ ~~~ ~~~ + \prod_{i=1}^L (1-\Pr[U_{i} \ge b_X - \epsilon_4 |X]) \\
	&= (1-\Pr[U \le a_X + \epsilon_4 |X])^L + (1-\Pr[U \ge b_X - \epsilon_4 | X])^L,
	\end{align*}
	where the last equality follows since agents are i.i.d.~conditioned on $X$. Since $f_{U|X}$ is continuous and does not vanish at $a_X, b_X$:
	\begin{align*}
	&\lim_{u \to a_X \textrm{ or } b_X} f_{U|X}(u|x) \ge \delta_U \\
	&\implies \Pr[U \le a_X + \epsilon_4 |X] \ge \delta_U \epsilon_4 ~~ \textrm{ and } \\
	&~~~ ~~~ ~~~ ~~~ ~~~ \Pr[U \ge b_X - \epsilon_4 |X] \ge \delta_U \epsilon_4.
	\end{align*}
	Therefore,
	\begin{align*}
	&\Pr[\mathcal{I}|X] \\
	&\le (1-\Pr[U \le a_X + \epsilon_4 |X])^L + (1-\Pr[U \ge b_X - \epsilon_4 |X])^L \\
	&\le 2(1-\delta_U \epsilon_4)^L,
	\end{align*}
	that is, the first term vanishes exponentially fast with $L$.
	
	Let us consider the second term of \eqref{eq:ach:error_bound}. Take random variables that denote cumulative distribution up to $U_{(1)}$ and tail distribution after $U_{(L)}$.
	\begin{align*}
	\eta &:= L \int_{a_X}^{U_{(1)}} f_{U|X}(u) du \ge \delta_U L (U_{(1)}-a_X), \\
	\xi &:= L \int_{U_{(L)}}^{b_X} f_{U|X}(u) du \ge \delta_U L (b_X-U_{(L)}),
	\end{align*}
	where $a_X \le U_{(1)} \le a_X+\epsilon_4, b_X-\epsilon_4 \le U_{(L)} \le b_X$. Then, the marginal and joint distributions are given in \cite{AkcayHL1996}
	\begin{align*}
	f_{\xi} (s) &= f_{\eta} (s) = \left( 1-\frac{s}{L}\right)^{L-1} ~~~ 0 \le s \le L, \\
	f_{\xi, \eta} (s_1, s_2) &= \frac{L-1}{L} \left( 1- \frac{s_1+s_2}{L}\right)^{L-2},
	\end{align*}
	for $s_1, s_2 \ge 0, s_1+s_2\le L$.
	As $L \to \infty$, $\xi$ and $\eta$ are asymptotically independent and $f_{\xi}(s), f_{\eta}(s) \to e^{-s}$. From the definition of $\xi, \eta$, 
	\begin{align*}
	&|a_X+b_X - (U_{(1)} + U_{(L)})|^r \\
	&= \left( (U_{(1)} - a_X) + (b_X - U_{(L)}) \right)^r \\
	&\le \frac{2^r (\xi^r + \eta^r)}{(L\delta_U)^r},
	\end{align*}
	where the last inequality follows from Prop.~\ref{prop:power_r_UB} and the definitions of $\eta$ and $\xi$. Therefore, when $L$ is large the second term is 
	\begin{align*}
	&\int_{\mathcal{I}^c} \left| a_X + b_X - U_{(1)} + U_{(L)} \right|^r \\
	&~~~ ~~~ ~~~ ~~~ ~~~ \times f_{u_{(1)}, u_{(L)}|X}(u_{(1)}, u_{(L)}|x) du_{(1)} du_{(L)} \\
	&\le \frac{2^r}{(L\delta_U)^r} \int_0^{z_1} \int_0^{z_2} (s_1^r + s_2^r) f_{\xi, \eta} (s_1, s_2) ds_1 ds_2,
 	\end{align*}
	with $z_1:=L(1-F_{U|X}(b_X-\epsilon_4))$ and $z_2:=LF_{U|X}(a_X+\epsilon_4)$. 
	
	Combining all of the above,
	\begin{align*}
	&\Rsum^r D(L,R) \\
	&\le \Rsum^r \left( \frac{K_U}{2} \right)^r \times \\
	&~~~ ~~\mathbb{E}_X \left[ \frac{2^r}{(L\delta_U)^r} \int_0^{z_1} \int_0^{z_2} (s_1^r + s_2^r) f_{\xi, \eta} (s_1, s_2) ds_1 ds_2 + \epsilon \right].
	\end{align*}
	When $L$ tends to infinity, using the fact that $\int_0^{\infty} s^r e^{-s} ds = r!$,
	\begin{align*}
	&\lim_{L \to \infty} \Rsum^r D(L,\Rsum) \\
	&\le \frac{K_U^r I(Y;U|X)^r }{\delta_U^r} \left( 2 \int_0^{\infty} s^r e^{-s} ds \right) \\
	&= 2 (r!) \left( \frac{K_U I(Y;U|X)}{\delta_U} \right)^r.
	\end{align*}
	Taking infimum over $\mathcal{S}_{\textsf{n-reg}}$ and decreasing the mutual information to zero complete the achievability proof.
\end{IEEEproof}

\subsection{Converse Coding Theorem}
The Chazan-Zakai-Ziv bound \cite{ChazanZZ1975} holds even under the non-regularity conditions (B1) and (B2) unlike the Cram\'{e}r-Rao lower bound. Hence, we will make use of the generalized Chazan-Zakai-Ziv bound to show the converse. The next lemma is a generalized version of the Chazan-Zakai-Ziv bound. The proof is an easy extension of special case $r=2$ \cite{ChazanZZ1975, BellSEV1997}, but for the sake of completeness, included in App.~\ref{app:CZZ_pf}.
\begin{lemma}[Chazan-Zakai-Ziv Bound for $r \in \mathbb{N}$] \label{lem:CZZ_abs}
	Consider $X \in [0, 1]$ and its noisy observation $Z$. Then,
	\begin{align*}
	\mathbb{E}\left[ |X - \hat{X}|^r \right] \ge \int_{h=0}^{1} r 2^{-r} h^{r-1} \int_{x=0}^{1-h} \phi(x, x+h) dx dh,
	\end{align*}
	where
	\begin{align*}
	\phi(x, x+h):=\frac{f_X(x) + f_X(x + h)}{2} P_{\textsf{min}} (x, x + h)
	\end{align*}
	and $P_{\textsf{min}}$ is the minimum probability of error of binary hypothesis testing with $H_0: Z \sim f_{Z|x}$ and $H_1: Z \sim f_{Z|x+h}$.
\end{lemma}
\begin{IEEEproof}
	See App.~\ref{app:CZZ_pf}.
\end{IEEEproof}

We suppose $X_{-t}^n$ is given at the moment of decoding and estimating $X(t)$, therefore $P_{\textsf{min}}$ in the above lemma is a function of $f_{C|X(t)}$ given $X_{-t}^n$. Recall that the same argument in Sec.~\ref{subsubsec:coding_rate_LB} gives the sum-rate lower bound
\begin{align*}
\Rsum \ge \frac{L}{n} \sum_{t=1}^n \mathbb{E}_{\breve{X}_t}[I(Y(t) ; U(t, \breve{X}_t) | X(t))].
\end{align*}
\begin{theorem}[Converse for Non-regular CEO Problem] \label{thm:non_reg_converse}
	For any $f_{U|Y}$ and $r \ge 1$, the following holds.
	\begin{align*}
	\beta_{\textsf{n-reg}} \ge \frac{r}{2^r} \int_{\tilde{h}=0}^1 \tilde{h}^{r-1} \int_{x=0}^{1} f_X(x) \theta(x, \tilde{h}) dx d\tilde{h},
	\end{align*}
	where
	\begin{align*}
	&\theta(x,\tilde{h}):=\lim_{a \to 0} \min_{ \substack{U:\\ X-Y-U \\ I(Y;U|X)=a }} I(Y;U|X)^r e^{-\tilde{h} g(x)}, \\
	&g(x)= \\
	&\frac{d}{d\Delta} \left( - \min_{s \in [0,1]} \log \left( \int f_{U|X}^s (u|x) f_{U|X}^{1-s} (u|x + \Delta) dy \right) \right) \Bigg\vert_{\Delta=0}.
	\end{align*}
\end{theorem}
\begin{IEEEproof}
	\begin{align*}
	&D^n(X^n, \hat{X}^n) \\
	&= \frac{1}{n} \sum_{t=1}^n \mathbb{E} \left[ |X(t) - \hat{X}(t)|^r \right] \\
	&\stackrel{(a)}{\ge} \frac{1}{n} \frac{r}{2^{r+1}} \sum_{t=1}^n \int_{h=0}^1 h^{r-1} \\
	&~~~ ~~~ ~~~ \times \int_{x=0}^{1-h} (f_X(x) + f_X(x+h)) P_{\textsf{min}} (x, x+h;t) dx dh \\
	&= \frac{1}{n} \frac{r}{2^{r+1}L^r} \sum_{t=1}^n \int_{h=0}^1 (Lh)^{r-1} \\
	&~~~ ~~~ \times \int_{x=0}^{1-h} (f_X(x) + f_X(x+h)) P_{\textsf{min}} (x, x+h;t) dx d(hL) \\
	&\stackrel{(b)}{=} \frac{1}{n} \frac{r}{2^{r+1}L^r} \sum_{t=1}^n \int_{\tilde{h}=0}^1 \tilde{h}^{r-1} \times \\
	&~~~  \int_{x=0}^{1-\frac{\tilde{h}}{L}} \left(f_X(x) + f_X(x+\frac{\tilde{h}}{L})\right) P_{\textsf{min}} \left(x, x+\frac{\tilde{h}}{L};t\right) dx d\tilde{h},
	\end{align*}
	where (a) follows from the Chazan-Zakai-Ziv bound in Lem.~\ref{lem:CZZ_abs}, and (b) is obtained by letting $\tilde{h} = hL$. Also, the Chernoff-Stein lemma \cite{CoverT1991} gives
	\begin{align*}
	P_{\textsf{min}}\left(x, x+\frac{\tilde{h}}{L} ; t \right) = e^{-LC(x, x+\frac{\tilde{h}}{L};t)},
	\end{align*}
	where $C\left(x, x+\frac{\tilde{h}}{L};t\right)$ is the Chernoff information between two conditional densities of codeword given $(x(t), x_{-t}^{n})$ and $(x(t) + \frac{\tilde{h}}{L}, x_{-t}^{n})$. Since $L \to \infty$, the quantity $G_{x}\left( \frac{\tilde{h}}{L};t \right):= C\left(x, x+\frac{\tilde{h}}{L};t\right)$ can be approximated by the Maclaurin expansion
	\begin{align*}
	G_{x}(\Delta;t) = G_{x}(0;t) + \Delta \cdot G_{x}'(0;t) + O(\Delta^2).
	\end{align*}
	As $G_{x}(0;t) = 0$, we have
	\begin{align*}
	P_{\textsf{min}}\left(x, x+\frac{\tilde{h}}{L};t\right) = e^{-LC(x, x+\frac{\tilde{h}}{L};t)} = e^{-\tilde{h} G_{x}'(0;t) + O(L^{-1})}.
	\end{align*}
	Therefore, for large $L$,
	\begin{align*}
	& D^n(X^n, \hat{X}^n) \\
	&\ge \frac{1}{n} \frac{r}{2^{r+1}L^r} \sum_{t=1}^n \int_{\tilde{h}=0}^1 \tilde{h}^{r-1} \times \\
	&~~~ ~~~ \int_{x=0}^{1-\frac{\tilde{h}}{L}} \left(f_X(x) + f_X(x+\frac{\tilde{h}}{L})\right) e^{-\tilde{h} G_{x}'(0;t) + O(L^{-1})} dx d\tilde{h} \\
	&\stackrel{(c)}{=} \frac{1}{n} \frac{r}{2^{r+1}L^r} \sum_{t=1}^n \int_{\tilde{h}=0}^1 \tilde{h}^{r-1} \int_{x=0}^{1} 2f_X(x) e^{-\tilde{h} g(x;t)} dx d\tilde{h} \\
	&= \frac{1}{n} \frac{r}{2^{r}L^r} \sum_{t=1}^n \int_{\tilde{h}=0}^1 \tilde{h}^{r-1} \int_{x=0}^{1} f_X(x) e^{-\tilde{h} g(x;t)} dx d\tilde{h},
	\end{align*}
	where (c) follows from the fact that $L$ is sufficiently large and $g(x;t)=G_{x}'(0;t)$, i.e.,
	\begin{align*}
	&g(x;t) = \frac{d}{d\Delta} \Big( - \min_{s \in [0,1]} \log \Big( \int f_{C|X(t), X_{-t}^n}^s (c|x, x_{-t}^n) \times \\
	&~~~ ~~~ ~~~ ~~~ ~~~ ~~~ ~~~ ~~~ f_{C|X(t), X_{-t}^n}^{1-s} (c|x + \Delta, x_{-t}^n) dy \Big) \Big) \Bigg\vert_{\Delta=0}.
	\end{align*}
	Then Jensen's inequality yields a further lower bound on $D^n(X^n, \hat{X}^n)$:
	\begin{align*}
	& D^n(X^n, \hat{X}^n) \\
	&\ge \frac{1}{n} \frac{r}{2^{r}L^r} \sum_{t=1}^n \int_{\tilde{h}=0}^1 \tilde{h}^{r-1} \int_{x=0}^{1} f_X(x) e^{-\tilde{h} g(x;t)} dx d\tilde{h} \\
	&=  \frac{r}{2^{r}L^r} \int_{\tilde{h}=0}^1 \tilde{h}^{r-1} \int_{x=0}^{1} f_X(x) \left( \frac{1}{n} \sum_{t=1}^n e^{-\tilde{h} g(x;t)} \right) dx d\tilde{h} \\
	&\ge \frac{r}{2^{r}L^r} \int_{\tilde{h}=0}^1 \tilde{h}^{r-1} \int_{x=0}^{1} f_X(x) e^{-\frac{1}{n} \sum_{t=1}^n \tilde{h} g(x;t)} dx d\tilde{h}.
	\end{align*}
	
	Consequently, we have \eqref{eq:long_eq} in the next page,
	\begin{figure*}[ht]
	\begin{align}
	&\Rsum^r D^n(X^n, \hat{X}^n) \nonumber \\
	&\ge \frac{r}{2^rL^r} \left( \frac{L}{n} \sum_{t=1}^n \mathbb{E}_{\breve{X}_t}[I(Y(t);U(t, \breve{X}_t) | X(t))] \right)^r \int_{\tilde{h}=0}^1 \tilde{h}^{r-1} \int_{x=0}^{1} f_X(x) e^{-\frac{1}{n} \sum_{t=1}^n \tilde{h} g(x;t)} dx d\tilde{h} \nonumber \\
	&= \frac{r}{2^r} \left( \frac{1}{n} \sum_{t=1}^n \mathbb{E}_{\breve{X}_t}[I(Y(t);U(t, \breve{X}_t) | X(t))] \right)^r \int_{\tilde{h}=0}^1 \tilde{h}^{r-1} \int_{x=0}^{1} f_X(x) e^{-\frac{1}{n} \sum_{t=1}^n \tilde{h} g(x;t)} dx d\tilde{h} \nonumber \\
	&\stackrel{(d)}{\ge} \frac{r}{2^r} \left( \prod_{t=1}^n \mathbb{E}_{\breve{X}_t}[I(Y(t);U(t, \breve{X}_t) | X(t))] \right)^{r/n} \int_{\tilde{h}=0}^1 \tilde{h}^{r-1} \int_{x=0}^{1} f_X(x) e^{-\frac{1}{n} \sum_{t=1}^n \tilde{h} g(x;t)} dx d\tilde{h} \nonumber \\
	&= \frac{r}{2^r} \left( \prod_{t=1}^n \mathbb{E}_{\breve{X}_t}[I(Y(t);U(t, \breve{X}_t) | X(t))] \right)^{r/n} \int_{\tilde{h}=0}^1 \tilde{h}^{r-1} \int_{x=0}^{1} f_X(x) \left( \prod_{t=1}^n e^{-\tilde{h} g(x;t)} \right)^{1/n} dx d\tilde{h} \nonumber \\
	&= \frac{r}{2^r} \int_{\tilde{h}=0}^1 \tilde{h}^{r-1} \int_{x=0}^{1} f_X(x) \left( \prod_{t=1}^n (\mathbb{E}_{\breve{X}_t}[I(Y(t);U(t, \breve{X}_t) | X(t))])^r e^{-\tilde{h} g(x;t)} \right)^{1/n} dx d\tilde{h} \nonumber \\
	&\ge \frac{r}{2^r} \int_{\tilde{h}=0}^1 \tilde{h}^{r-1} \int_{x=0}^{1} f_X(x) \left( \min_{t, \breve{X}_t} I(Y(t);U(t, \breve{X}_t) | X(t))^r e^{-\tilde{h} g(x;t)} \right) dx d\tilde{h}, \label{eq:long_eq}
	\end{align}
	\end{figure*}
	where (d) follows from the arithmetic mean-geometric mean inequality. Hence, changing $(C, \breve{X}_t)$ to $U$ yields the final form:
	\begin{align*}
	&\Rsum^r D^n(X^n, \hat{X}^n) \\
	&\ge \frac{r}{2^r} \int_{\tilde{h}=0}^1 \tilde{h}^{r-1} \int_{x=0}^{1} f_X(x) \left( \min_{U} I(Y;U|X)^r e^{-\tilde{h} g(x)} \right) dx d\tilde{h},
	\end{align*}
	where
	\begin{align*}
	&g(x) = \\
	&\frac{d}{d\Delta} \left( - \min_{s \in [0,1]} \log \left( \int f_{U|X}^s (u|x) f_{U|X}^{1-s} (u|x + \Delta) dy \right) \right) \Bigg\vert_{\Delta=0}.
	\end{align*}
	Limiting $I(Y;U|X) \to 0$ proves the converse.
\end{IEEEproof}

\section{Equivalence of Quadratic and Logarithmic Distortions} \label{sec:equivalence}
In this section, we show that quadratic distortion $D_{\textsf{Q}}$ and logarithmic distortion $D_{\textsf{Log}}$ \cite{CourtadeW2014} are in fact asymptotically (in number of agents) equivalent under some conditions. Suppose we have a set of observations $Z$ and the source $X$ to be estimated. Then, it is known that $D_{\textsf{Q}} = \mathbb{E}[(X-\hat{X}(Z))^2] \ge \textsf{Var}(X|Z)$ and $D_{\textsf{Log}} \ge h(X|Z)$ \cite{CourtadeW2014}, and the equalities are in fact attainable. In addition, those two are in general related by the entropy power inequality \cite{CoverT1991},
\begin{align*}
\textsf{Var}(X|Z) \ge \tfrac{1}{2\pi e} e^{2h(X|Z)} \implies D_{\textsf{Q}} \ge \tfrac{1}{2\pi e} e^{2D_{\textsf{Log}}}.
\end{align*}
We previously showed that the equality holds in the case of the jointly Gaussian CEO problem with finite number of agents \cite{SeoV2016} due to the entropy maximization property of the Gaussian distribution. Here we extend it to our regular CEO problem and provide conditions for which $D_{\textsf{Q}}$ and $D_{\textsf{Log}}$ are asymptotically equivalent under the entropy power conversion $D_{\textsf{Q}} = \tfrac{1}{2\pi e} 2^{D_{\textsf{Log}}}$ when $L \to \infty$.

To argue the equivalence, we restate that each agent's test channel (thus, subsequent encoding scheme) $f_{U_i|Y_i}$ is identical to $f_{U|Y}$ as assumed in the previous sections. Also, beyond the regular model in Sec.~\ref{sec:regular_ceo}, we further suppose the following conditions on test channels for logarithmic optimal codewords \cite{Vaart1998, DasGupta2008}:
\begin{itemize}	
\item[(C1)] For all $x \in \mathcal{X}$, it holds that $\int_{\mathcal{U}} \tfrac{\partial^2}{\partial x^2} f_{U|X}(u|x) du = 0$. In addition, the Fisher information is finite and positive, i.e., 	$0 < I_U(x):=\mathbb{E}_{U|x}[(\tfrac{\partial}{\partial x} \log f_{U|X}(U|x))^2] < \infty$ for all $x \in \mathcal{X}$.

\item[(C2)] Let $x_0$ denote the true source. Then, there exists $k(u)$ such that $| \tfrac{\partial^2}{\partial x^2} f_{U|x}(u|x) | \le k(u)$ on small neighborhood of $x_0$ and such that $\mathbb{E}_{x_0}[k(U)]$ is finite.
\end{itemize}
Define $\mathcal{S}_{\textsf{reg}}'$ to be the set of $U$s that satisfy (C1) and (C2) as well as (A5). Note that although $\mathcal{S}_{\textsf{reg}}' \subset \mathcal{S}_{\textsf{reg}}$, it only affects the constant factor in Thm.~\ref{thm:regular_ceo_ach}.

Before proceeding to the equivalence, let us state the Bernstein-von Mises theorem which is often referred to as \emph{asymptotic normality of posterior}, without the prior having an effect.
\begin{lemma}[Bernstein-von Mises \cite{Vaart1998, DasGupta2008}]
Suppose (C1) and (C2) as well as (A1)--(A5) hold. Then, for any $x_0 \in \mathcal{X}$,
\begin{align*}
d_{\textsf{TV}}( f(X|U^L), \mathcal{N}(\hat{X}_{\textsf{MLE}}, (LI_U(x_0))^{-1}) ) \to 0
\end{align*}
as $L \to \infty$ with $f_{Y|x_0} \textrm{-probability } 1$, where $\hat{X}_{\textsf{MLE}}$ and $d_{\textsf{TV}}$ denote the maximum likelihood estimator and total variational distance, respectively.
\end{lemma}

Now we can prove the equivalence, which relies on the Bernstein-von Mises theorem.
\begin{theorem}\label{thm:equivalence}
Suppose that the optimal codebook for logarithmic distortion is generated from a member of $\mathcal{S}_{\textsf{reg}}'$. Then, $D_{\textsf{Q}}(L, \Rsum)$ and $D_{\textsf{Log}}(L, \Rsum)$ asymptotically (in number of agents) satisfy
\begin{align*}
D_{\textsf{Log}}(L, \Rsum) - \frac{1}{2} \log \left( 2 \pi e D_{\textsf{Q}}(L, \Rsum) \right) \to 0
\end{align*}
as $L, \Rsum \to \infty$.
\end{theorem}
\begin{IEEEproof}
Let us consider the quadratic optimal codebook and fix some codewords $(w_1, w_2, \ldots, w_L)$. Then, incurred quadratic distortion is
\begin{align*}
&\left(D_{\textsf{Q}}^n(L, \Rsum | \{w_i\}_{i=1}^L) \right)^n \\
&= \left( \frac{1}{n} \sum_{t=1}^{n} \mathbb{E}\left[ |X(t) - \hat{X}(t)|^2 \Big| \{w_i\}_{i=1}^L \right] \right)^n \\
&= \left( \frac{1}{n} \sum_{t=1}^{n} \textsf{Var}(X(t)|\{w_i\}_{i=1}^L) \right)^n \\
&\stackrel{(a)}{\geq} \prod_{t=1}^{n} \textsf{Var} (X(t)|\{w_i\}_{i=1}^L) \\
&\stackrel{(b)}{\geq} \prod_{t=1}^{n} \frac{1}{2 \pi e} e^{2h(X(t)|\{w_i\}_{i=1}^L)} \\
&= \frac{1}{(2 \pi e)^n} e^{2 \sum h(X(t)|\{w_i\}_{i=1}^L)} \\
&\stackrel{(c)}{=} \frac{1}{(2 \pi e)^n} e^{2nD_{\textsf{Log}}^n(L, \Rsum | \{w_i\}_{i=1}^L)} \\
&= \left( \frac{1}{2 \pi e} e^{2D_{\textsf{Log}}^n(L, \Rsum | \{w_i\}_{i=1}^L)} \right)^n,
\end{align*}
where (a) follows from the arithmetic-geometric inequality; (b) follows from the fact that the Gaussian distribution maximizes differential entropy for a given variance; and (c) follows by declaring the true posterior distribution \cite[Lem.~1]{CourtadeW2014}. Hence, taking an expectation over all codewords,
\begin{align*}
&D_{\textsf{Q}}^n(L, \Rsum) \\
&= \mathbb{E} \left[ D_{\textsf{Q}}^n(L, \Rsum | \{W_i\}_{i=1}^L) \right] \\
&\ge \mathbb{E}\left[ \frac{1}{2 \pi e} e^{2D_{\textsf{Log}}^n(L, \Rsum | \{W_i\}_{i=1}^L)} \right] \\
&\stackrel{(d)}{\ge} \frac{1}{2 \pi e} e^{2 \mathbb{E}[D_{\textsf{Log}}^n(L, \Rsum | \{W_i\}_{i=1}^L)]} \\
&= \frac{1}{2 \pi e} e^{2 \tilde{D}_{\textsf{Log}}^n(L, \Rsum)},
\end{align*}
where (d) follows from Jensen's inequality and $\tilde{D}_{\textsf{Log}}^n$ is the logarithmic distortion incurred by the optimal codebook in the quadratic case. It is therefore obvious that the logarithmic optimal codebook achieves a smaller distortion. This concludes one direction
\begin{align*}
D_{\textsf{Q}}^n(L, \Rsum) \ge \frac{1}{2 \pi e} e^{2 D_{\textsf{Log}}^n (L, \Rsum)}.
\end{align*}

To show the other direction, consider the optimal codebook for the logarithmic distortion.
\begin{align*}
&D_{\textsf{Log}}^n (L, \Rsum) = h(X|U^L) \\
&\stackrel{(a)}{=} \frac{1}{2} \log \left( 2 \pi e \textsf{Var}(X|U^L) \right) \\
&\stackrel{(b)}{\ge} \frac{1}{2} \log \left( 2 \pi e D_{\textsf{Q}}^n(L, \Rsum) \right),
\end{align*}
where (a) is in fact `$\ge$', but the equality asymptotically holds by the Bernstein-von Mises theorem, and (b) follows since the optimal codebook in the logarithmic case is suboptimal for the quadratic distortion. This shows the other direction and the theorem is proved.
\end{IEEEproof}

\section{Discussion} \label{sec:discussion}
We studied two continuous-alphabet CEO problems---regular and non-regular---and found sum-rate asymptotics $\mathcal{O}(\Rsum^{-r/2})$ and $\mathcal{O}(\Rsum^{-r})$ respectively for $|x-\hat{x}|^r$ distortion. For the regular CEO problem, we used a practical sample median to estimate the source. For the non-regular CEO problem, we used a midrange estimator that relies heavily on the non-vanishing property of the conditional distribution at boundaries. Converse bounds are also provided by the Shannon lower bound and the Chazan-Zakai-Ziv bound respectively, but whether or not they are tight is generally unknown.

Our achievable schemes leave several interesting future directions. First note that the median estimator may not be the best, but achieves the correct sum-rate asymptotics, at least for Gaussian. For instance, the (scaled version of) sample mean estimator in \cite{ViswanathanB1997} turns out to be the best estimator for the quadratic Gaussian CEO problem even in the non-asymptotic regime \cite{Oohama2005, PrabhakaranTR2004} because the minimum mean-squared error estimator (MMSE) is in fact linear sum of codewords from the additive Gaussian test channel. To illustrate the pros and cons of the two estimators, consider a simple estimation problem of $X$ from observation $Y_i=X+Z_i, i \in [1:L]$ without coding, where $Y_i$ is a given observation, $Z_i$ is i.i.d.~drawn from some $f_Z$ with zero mean and $\sigma_Z^2$ variance. In this case, the sample mean estimator is distributed approximately $\mathcal{N}(X, \sigma_Z^2/L)$ by the central limit theorem so yields approximately $\sigma_Z^2/L$ quadratic distortion. However, the quadratic distortion induced by the median estimator is $(4Lf_Z^2(0))^{-1}$ according to Lem.~\ref{lem:Median_Gaussianity}. Since the performance of the median estimator is independent of $\sigma_Z^2$, the median estimator is more efficient when $Z$ is sufficiently heavy-tailed. So one may wonder under what conditions median estimation outperforms mean estimation in our asymptotic regime. However, it is not obvious as $I(Y;U|X)/\alpha_U^2$ is generally hard to evaluate.

The median estimation is also widely used for robust estimation. Even when only a couple of malicious agents corrupt their observations, the mean estimator fails to recover the population mean, whereas the median estimator still recovers the population median within a small error \cite{Huber1981}. Hence, it is interesting to consider an adversarial CEO problem like \cite{KosutT2008a}.

The fact that the median and mean estimators both achieve the same distortion decay at least for the Gaussian case implies that a broader class of estimators may achieve the same asymptotics. Thus, considering another type (or a class) of estimators is also interesting. Note that Lem.~\ref{lem:Median_Gaussianity} which is due to Gaussianity further suggests that a broader class of estimators called \emph{Consistent and Asymptotic Normal (CAN)} estimators perhaps achieves the same asymptotics; for example, the maximum likelihood estimator (MLE) is also CAN and furthermore asymptotically \emph{efficient} \cite{Wasserman2013}. The asymptotic normality of the MLE by the Bernstein-von Mises theorem is also a stepping stone to the equivalence of quadratic and logarithmic distortions in Sec.~\ref{sec:equivalence}. Since the Bernstein-von Mises theorem relies on a strong assumption that the $\{U_i\}$ are conditionally independent and identically distributed, it is natural to investigate relaxing the identicality and independence for more realistic and general models.

For the non-regular CEO problem, we used the midrange estimator that relies heavily on the non-vanishing property of conditional distribution at boundaries. However, note that the Chazan-Zakai-Ziv bound does not require such a non-vanishing condition. So, it would be an interesting future direction to find a generalization of the non-regular model and/or its matching estimation scheme in (non-)asymptotic regime. Moreover, since (A1)--(A4) and (B1)--(B2) do not form a disjoint partition, there are other models that do not belong to either of the two. For example, when the observational noise is additive triangular, it does not satisfy non-vanishing probability density in (B2) so the midrange estimator does not work. We do not know its distortion decay, which would be of interest to determine.

As we have seen, the regular model attains $\mathcal{O}(\Rsum^{-r/2})$ distortion decay. In addition, it is known that the jointly Gaussian model, a member of the regular model class, is the least compressible model among all finite-variance additive noise models, i.e., it admits the largest distortion for given compression rates \cite{AsnaniSAW2015}. Therefore, we can conclude that the regular model is the hardest model to estimate in the distortion decay sense. In contrast, non-regular models that have $\mathcal{O}(\Rsum^{-r})$ decay are easier to estimate than the regular model. Extending this argument, it is interesting to classify various models according to their distortion decay.

\appendices
\section{Proof of Lem.~\ref{lem:Median_Gaussianity}: Absolute Central Moments for Sample Median} \label{app:pf_of_med_gau}
Let us introduce some notations first. Let $W$ be the Gaussian random variable with distribution $\mathcal{N}\left(\textsf{med}(V), \frac{1}{4 L f^2(\textsf{med}(V))} \right)$. Let $\xi_m := \mathbb{E}[V_{(m+1)}]$, and notice that $\xi_m \neq \textsf{med}(V)$ in general since a sample median, $V_{(m+1)}$, is biased unless $V$ is unskewed. Also, let $\gamma_r, \gamma_r'$ be the absolute central moments of $V_{(m+1)}$ and $W$ of order $r$, i.e.,
\begin{align*}
\gamma_r &:= \mathbb{E}[|V_{(m+1)}-\xi_m|^r], \\
\gamma_r' &:= \mathbb{E}[|W - \textsf{med}(V)|^r],
\end{align*}
and $\rho_r, \rho_r'$ be central moments of $V_{(m+1)}$ and $W$, i.e.,
\begin{align*}
\rho_r &:= \mathbb{E}[(V_{(m+1)}-\xi_m)^r], \\
\rho_r' &:= \mathbb{E}[(W - \textsf{med}(V))^r].
\end{align*}
Our proof is then based on the following result.
\begin{proposition}[\cite{ChuH1955}] \label{prop:ChuH1955}
	For all $r \ge 2$, $\lim_{m \to \infty} \rho_r = \rho_r'$ provided that all quantities exist and finite, and $\lim_{m \to \infty} \xi_m = \textsf{med}(V)$.
\end{proposition}

Due to the triangle inequality and Prop.~\ref{prop:power_r_UB},
\begin{align*}
&\mathbb{E}[ |V_{(m+1)} - \textsf{med}(V)|^r ] \\
&= \mathbb{E}[ |V_{(m+1)} - \xi_m + \xi_m - \textsf{med}(V)|^r ] \\
&\le \mathbb{E}[|V_{(m+1)} - \xi_m|^r] + \epsilon,
\end{align*}
where the last equality follows since $\xi_m, \textsf{med}(V)$ are deterministic quantities. Furthermore, due to  Prop.~\ref{prop:ChuH1955}, we can take large $m$ for any positive $\delta$ such that $|\xi_m - \textsf{med}(V)|^r \le \delta$.

Consider the first term. Since $\rho_r \to \rho_r'$, we know that $\gamma_r, \gamma_r'$ are also bounded. Letting $A:=V_{(m+1)} - \xi_m$ and $B:=W - \textsf{med}(V)$ for brevity, take a large $p \in \mathbb{N}$ such that
\begin{align*}
\Big|\mathbb{E}[|A|^r] - \mathbb{E}[|A|^r \wedge p]\Big| \le \delta ~ \textrm{ and } ~ \Big|\mathbb{E}[|B|^r] - \mathbb{E}[|B|^r \wedge p]\Big| \le \delta,
\end{align*}
where $\wedge$ is the minimum operator. Then,
\begin{align*}
&\Big| \mathbb{E}[|A|^r] - \mathbb{E}[|B|^r] \Big| \\
&= \Big| \mathbb{E}[|A|^r - |A|^r \wedge p + |A|^r \wedge p] \\
&~~~ ~~~ ~~~ ~~~ ~~~ - \mathbb{E}[|B|^r - |B|^r \wedge p + |B|^r \wedge p] \Big| \\
&\le 2 \delta + \Big|\mathbb{E}[|A|^r \wedge p] - \mathbb{E}[|B|^r \wedge p]\Big|.
\end{align*}
Note that $|\cdot|^r \wedge p$ is a bounded continuous function and $A \to B$ in distribution as $m \to \infty$ by Lem.~\ref{lem:Median_Gaussianity}. Therefore, we can take large $m$ such that $\Big|\mathbb{E}[|A|^r \wedge p] - \mathbb{E}[|B|^r \wedge p]\Big| \le \delta$ by the continuous mapping theorem, which leads us to
\begin{align*}
\Big| \mathbb{E}[|A|^r] - \mathbb{E}[|B|^r] \Big| \le 3 \delta.
\end{align*}
Hence, the first term satisfies
\begin{align*}
\mathbb{E}[ |V_{(m+1)} - \xi_m|^r ] \le \mathbb{E}[|B|^r] + 3 \delta + \delta.
\end{align*}
Note that $\mathbb{E}[|B|^r]$ is the $r$th absolute central moment of Gaussian. For $Z \sim \mathcal{N}(\mu, \sigma^2)$, it holds that $\mathbb{E}\left[ |X-\mu|^r \right] = (2\sigma^2)^{r/2} \frac{\Gamma(\frac{r+1}{2})}{\sqrt{\pi}}$. Hence,
\begin{align*}
\mathbb{E}[|B|^r] = \left( \frac{1}{2Lf^2(\textsf{med}(V))} \right)^{r/2} \frac{\Gamma(\frac{r+1}{2})}{\sqrt{\pi}},
\end{align*}
which completes the proof since $\delta$ is arbitrary.

\section{Distortion Bounds in Achievability} \label{app:pf_of_epsilon}
The next proposition is a part of the proof of the regular model achievability.
\begin{proposition}[Regular CEO Problem] \label{prop:regular_epsilon}
\begin{align*}
&K_U^r \mathbb{E} \left[ |\textsf{med}(U|X) - U_{(m+1)}(t) + U_{(m+1)}(t) - \hat{U}_{(m+1)}(t)|^r \right] \\
&\le K_U^r \mathbb{E} \left[ |\textsf{med}(U|X) - U_{(m+1)}(t)|^r \right] + \epsilon_1.
\end{align*}
\end{proposition}
\begin{IEEEproof}
For the sake of brevity, define $A:= |\textsf{med}(U|X) - U_{(m+1)}(t)|$ and $B:= |U_{(m+1)}(t) - \hat{U}_{(m+1)}(t)|$. By the triangle inequality,
\begin{align*}
&K_U^r \mathbb{E} \left[ |\textsf{med}(U|X) - U_{(m+1)}(t) + U_{(m+1)}(t) - \hat{U}_{(m+1)}(t)|^r \right] \\
&\le K_U^r \mathbb{E} \left[ (A+B)^r \right] \\
&= K_U^r \mathbb{E} \left[ A^r + \binom{r}{1} A^{r-1}B + \cdots + \binom{r}{r-1} AB^{r-1} + B^r \right].
\end{align*}
Due to the Cauchy-Schwartz inequality that $\mathbb{E}[A^i B^j] \le \sqrt{ \mathbb{E}[A^{2i}] \mathbb{E}[B^{2j}]}$ for $j \ge 1$, it is sufficient to show that $\mathbb{E}[B^{2j}] \le \epsilon$ for all $1 \le j \le r$. Hence,
\begin{align}
&\mathbb{E}[B^{2j}] = \mathbb{E} \left[ |U_{(m+1)} - \hat{U}_{(m+1)}|^{2j} \right] \label{eq:error_eq_to_show} \\
&= \mathbb{E} \left[ |U_{(m+1)} - q(U_{(m+1)}) + q(U_{(m+1)}) - \hat{U}_{(m+1)}|^{2j} \right] \nonumber \\
&\le 2^{2j} \mathbb{E} \left[ |U_{(m+1)} - q(U_{(m+1)}) |^{2j} \right] \nonumber \\
&~~~ ~~~ ~~~ ~~~ ~~~ + 2^{2j} \mathbb{E} \left[ |q(U_{(m+1)}) - \hat{U}_{(m+1)} |^{2j} \right] \nonumber \\
&\le 2^{2j} \mathbb{E} \left[ |U_{(m+1)} - q(U_{(m+1)}) |^{2j} \right] \nonumber \\
&~~~ ~~~ ~~~ ~~~ ~~~ + 2^{4j} \mathbb{E} \left[ |q(U_{(m+1)}) - \textsf{med}(\{\tilde{U}_i\}_{i=1}^L) |^{2j} \right] \nonumber \\
&~~~ ~~~ ~~~ ~~~ ~~~ + 2^{4j} \mathbb{E} \left[ |\textsf{med}(\{\tilde{U}_i\}_{i=1}^L) - \hat{U}_{(m+1)} |^{2j} \right], \nonumber
\end{align}
where both inequalities are due to the triangle inequality and Prop.~\ref{prop:power_r_UB}.

Now the first and second terms can be small enough due to \eqref{eq:quant_assumption1}, i.e.,
\begin{align*}
2^{2j} \mathbb{E} \left[ |U_{(m+1)} - q(U_{(m+1)}) |^{2j} \right] &\le 2^{2j} \delta_0, \\
2^{4j} \mathbb{E} \left[ |q(U_{(m+1)}) - \textsf{med}(\{\tilde{U}_i\}_{i=1}^L) |^{2j} \right] &\le 2^{4j} \delta_0.
\end{align*}

Note that $\tilde{U}_i, \hat{U}_{(m+1)} \in \tilde{\mathcal{U}}$, a finite set. Hence, the last term can be also bounded using the Slepian-Wolf decoding error $\mathcal{B}$ in Prop.~\ref{prop:SW_decoding}, 
\begin{align*}
&2^{4j} \mathbb{E} \left[ |\textsf{med}(\{\tilde{U}_i\}_{i=1}^L) - \hat{U}_{(m+1)} |^{2j} \right] \\
&\le 2^{4j} (2 \tilde{u}_{\textsf{max}})^{2j} \Pr[\mathcal{B}] \le 2^{4j} (2 \tilde{u}_{\textsf{max}})^{2j} \lambda,
\end{align*}
and decreasing $\lambda \to 0$ with $n \to \infty$. Therefore, we can choose small $\delta$ and large $n$ appropriately so that \eqref{eq:error_eq_to_show} is bounded by $\epsilon_2 > 0$. This completes the claim.
\end{IEEEproof}

The next proposition is a part of the proof of the non-regular model achievability.
\begin{proposition}[Non-regular CEO Problem] \label{prop:non-regular_epsilon}
	For any $\epsilon > 0$, there exist a quantization scheme and block length $n$ such that
	\begin{align*}
	&\left(\frac{K}{2}\right)^r \mathbb{E} \Big[ \Big| |(a_X + b_X) - (U_{(1)} + U_{(L)})| \\
	&~~~ ~~~ ~~~ ~~~ ~~~ ~~~ + |(U_{(1)} + U_{(L)}) - (\hat{U}_{(1)} + \hat{U}_{(L)})| \Big|^r \Big] \\
	&\le \left(\frac{K}{2}\right)^r \mathbb{E} \left[ \left|(a_X + b_X) - (U_{(1)} + U_{(L)}) \right|^r \right] + \epsilon_3,
	\end{align*}
\end{proposition}
\begin{IEEEproof}
	For brevity, define $A:= |a_X + b_X) - (U_{(1)} + U_{(L)})|$ and $B:=|(U_{(1)} + U_{(L)}) - (\hat{U}_{(1)} + \hat{U}_{(L)})|$. Then we can write that
	\begin{align}
	& \left(\frac{K}{2}\right)^r \mathbb{E} \left[ (A+B)^r \right] \nonumber \\
	&= \left(\frac{K}{2}\right)^r \mathbb{E} \left[ A^r + \binom{r}{1} A^{r-1}B + \cdots + B^r \right]. \label{eq:non_regular_to_show}
	\end{align}
	Due to the Cauchy-Schwartz inequality that $\mathbb{E}[A^i B^j] \le \sqrt{ \mathbb{E}[A^{2i}] \mathbb{E}[B^{2j}]}$ for $i,j \ge 1$, it is sufficient to show that $\mathbb{E}[B^{2j}] \le \epsilon$ for all $1 \le j \le r$. Hence,
	\begin{align*}
	& \mathbb{E}[B^{2j}] = \mathbb{E} \left[ |(U_{(1)} + U_{(L)}) - (\hat{U}_{(1)} + \hat{U}_{(L)})|^{2j} \right] \\
	&= \mathbb{E} \Big[ |(U_{(1)} + U_{(L)}) - ( \tilde{U}_{(1)} + \tilde{U}_{(L)} ) \\
	&~~~ ~~~ ~~~ ~~~ ~~~  + ( \tilde{U}_{(1)} + \tilde{U}_{(L)} ) - (\hat{U}_{(1)} + \hat{U}_{(L)})|^{2j} \Big] \\
	&\le 2^{2j} \mathbb{E} \left[ |(U_{(1)} + U_{(L)}) - ( \tilde{U}_{(1)} + \tilde{U}_{(L)} )|^{2j} \right] \\
	&~~~ ~~~ ~~~ ~~~ + 2^{2j} \mathbb{E}\left[ |( \tilde{U}_{(1)} + \tilde{U}_{(L)}) - (\hat{U}_{(1)} + \hat{U}_{(L)})|^{2j} \right]
	\end{align*}
	
	The first term decomposes into two terms by Prop.~\ref{prop:power_r_UB} and taking sufficiently fine quantization points \eqref{eq:non_reg_quan},
	\begin{align*}
	& 2^{2j} \mathbb{E} \left[ |(U_{(1)} + U_{(L)}) - ( \tilde{U}_{(1)} + \tilde{U}_{(L)} )|^{2j} \right] \\
	&\le 2^{4j} \mathbb{E} \left[ \left| U_{(1)} - \tilde{U}_{(1)} \right|^{2j} \right] + 2^{4j} \mathbb{E}\left[\left|U_{(L)} - \tilde{U}_{(L)} \right|^{2j} \right] \\
	&\le 2^{4j+1} \delta_0.
	\end{align*}	
	
	To bound the second term, recall the decoding error probability $\Pr[\mathcal{B}] \le \lambda$ given in Prop.~\ref{prop:SW_decoding}. Then, applying Prop.~\ref{prop:power_r_UB},
	\begin{align*}
	&2^{2j} \mathbb{E} \left[ \left| \tilde{U}_{(1)} + \tilde{U}_{(L)} - \hat{U}_{(1)} - \hat{U}_{(L)} \right|^{2j} \right] \\
	&\le 2^{4j} \left( \mathbb{E}[|\tilde{U}_{(1)} - \hat{U}_{(1)}|^{2j}] + \mathbb{E}[|\tilde{U}_{(L)} - \hat{U}_{(L)}|^{2j}] \right) \\
	&\le 2^{4j} 2 (2 \tilde{u}_{\textsf{max}})^{2j} \Pr[\mathcal{B}] \le 2^{4j+1} (2 \tilde{u}_{\textsf{max}})^r \lambda,
	\end{align*}
	where $\tilde{u}_{\textsf{max}} := \max\{|\tilde{u}|: \tilde{u} \in \mathcal{U} \} < 1$ as $\mathcal{U} = [0,1]$.
	
	Hence, taking sufficiently fine quantization and taking sufficiently large $n$, we can bound \eqref{eq:non_regular_to_show} by any $\epsilon_3 > 0$.	
\end{IEEEproof}

\section{Proof of Prop.~\ref{prop:collection_delta}} \label{app:pf_collection_delta}
\begin{figure}[ht]
	\centering
	\includegraphics[width=.35\textwidth]{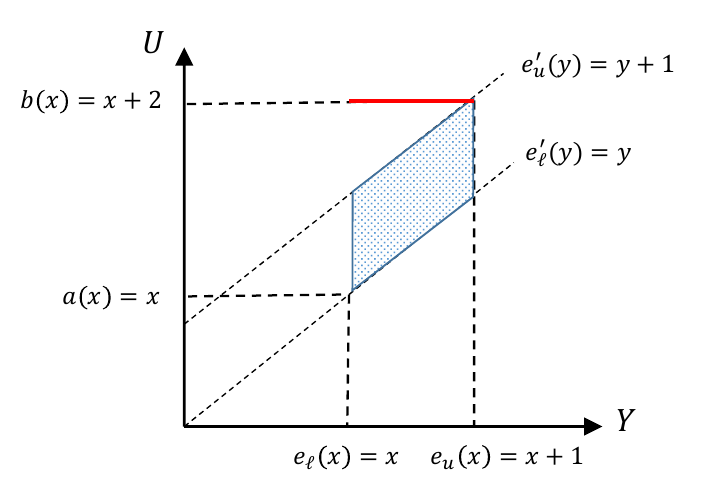}
	\caption{Illustration of $f_{Y, U| x}$ for $U=x+N+V$ where $N,V$ are uniformly distributed on $[0,1]$. The joint distribution is positive only on the shaded region.  }
	\label{fig:joint_pdf}
\end{figure}

Fix $x$, and suppose the support of $f_{Y|x}$ is $\mathcal{Y}_x = [e_{\ell}(x), e_u(x)]$, an interval. The case when $\mathcal{Y}_x \subsetneq [e_{\ell}(x), e_u(x)]$ (e.g., union of disjoint intervals) will be discussed later. By (B2), we know that $f_{Y|x}(e_{\ell}(x)), f_{Y|x}(e_{u}(x)) > 0$. Let $\mathcal{V} \subset [v_{\ell}, v_u]$ be the support of $f_V$. Even when $f_V$ is zero inside of $[v_{\ell}, v_u]$, it does not affect the support computation so we can assume $\mathcal{V} = [v_{\ell}, v_u]$ without loss of generality. As $V$ is independent and additive, the range of $Y+V$ is $[Y+v_{\ell}, Y+v_u]$ so $\mathcal{U}_x = [a(x), b(x)] = \cup_{y \in \mathcal{Y}_x} [y+v_{\ell}, y+v_u]$. For instance, when $U=X+N+V$ where $X, N, V$ are independently drawn from $\textsf{unif}[0,1]$, $\mathcal{U}_x = \cup_{y \in [x, x+1]} [y+0, y+1] = [x, x+2]$. This is illustrated in Fig.~\ref{fig:joint_pdf}.

As $f_{Y|x}$ is positive on $\mathcal{Y}_x$, we can take $0 < c < C$ such that $c < f_{Y|x}(y) < C$ on $\mathcal{Y}_x$. Then, by the marginalization law,
\begin{align*}
&f_{U|x}(b(x)) = \int_{y \in \mathcal{Y}_x} f_{U|Y}(b(x)|y) f_{Y|X}(y|x) dy \\
\implies& c \int_{y \in [e_{\ell}, e_u]} f_{U|Y}(b(x) |y) dy \\
& ~~~ ~~~ \le f_{U|x}(b(x)|x) \le C \int_{y \in [e_{\ell}, e_u]} f_{U|Y}(b(x) |y) dy.
\end{align*}
It is therefore equivalent to showing that $\int f_{U|Y}(b(x)|y) dy > \delta$ for some $\delta > 0$. However, $f_{U|Y}(b(x)|y)$ is positive only when $y=e_u(x)$. In other words, the set of points on which $f_{U|Y}(b(x)|y) > 0$ is Lebesgue measure zero. This in turn implies $\int f_{U|Y}(b(x)|y) dy = 0$. In the above example, this integration is the Riemann sum along the solid red line in Fig.~\ref{fig:joint_pdf}.

Next consider the case when $f_{Y|x}(y) = 0$ on some interior subsets of $[e_{\ell}(x), e_u(x)]$. That is, $\mathcal{Y}_x \subsetneq [e_{\ell}(x), e_u(x)]$. Take $g(y)$ such that $g(y) \ge f_{Y|x}(y)$ and $c < g(y) < C$ for all $y \in [e_{\ell}(x), e_u(x)]$. Then, we have
\begin{align*}
f_{U|x}(b(x)) &= \int_{y \in [e_{\ell}, e_u]} f_{U|Y}(b(x)|y) f_{Y|X}(y|x) dy \\
&\le \int_{y \in [e_{\ell}, e_u]} f_{U|Y}(b(x)|y) g(y) dy.
\end{align*}
Since $g(y)$ might not be a valid probability distribution, we have to normalize properly and obtain a valid distribution $\tilde{g}(y)$, but this does not change whether the right side is positive or not. Applying the above argument concludes that $f_{U|x}(b(x))$ cannot be positive in this case as well.

\section{Proof of Lem.~\ref{lem:CZZ_abs}: Chazan-Zakai-Ziv Bound for $r$th Power} \label{app:CZZ_pf}
Let us start with the following identity for a non-negative random variable $A$:
\begin{align*}
\mathbb{E}[A] = \int_{0}^{\infty} \Pr[A \ge t] dt.
\end{align*}

Letting $A = |\hat{X} - X|^r$ and $t = \left(\frac{h}{2}\right)^r$, we have the following identity by change of variables.
\begin{align*}
&\mathbb{E}[|X-\hat{X}|^r] \\
&= \int_{0}^{\infty} \Pr[|X-\hat{X}|^r \ge t] dt \\
&= \int_{0}^{\infty} r 2^{-r} h^{r-1} \Pr\left[|X-\hat{X}|^r \ge \left(\frac{h}{2}\right)^r \right] dh \\
&= \int_{0}^{\infty} r 2^{-r} h^{r-1} \Pr\left[|X-\hat{X}| \ge \frac{h}{2} \right] dh.
\end{align*}

Let us derive a lower bound on $\Pr \left[ |\hat{X}-X| \ge \frac{h}{2} \right]$.
\begin{align*}
&\Pr \left[ |\hat{X}-X| \ge \frac{h}{2} \right] \\
&= \Pr \left[ \hat{X} -X \ge \frac{h}{2} \right] + \Pr \left[ \hat{X} -X < -\frac{h}{2} \right] \\
&= \int_{0}^1 f_X(x) \Pr \left[ \hat{X} - X \ge \frac{h}{2} \Big| X = x \right] dx \\
&~~~ ~~~ ~~~ ~~~ + \int_{0}^1 f_X(x) \Pr \left[ \hat{X} - X < -\frac{h}{2} \Big| X = x \right] dx.
\end{align*}

By change of variable $x = t+h$ in the second integration, we have \eqref{eq:long_eq2} in the next page.
\begin{figure*}[t]
\begin{align}
&\Pr \left[ |\hat{X}-X| \ge \frac{h}{2} \right] \nonumber \\
&= \int_{0}^1 f_X(x) \Pr \left[ \hat{X} - X \ge \frac{h}{2} \Big| X = x \right] dx + \int_{-h}^{1-h} f_X(x) \Pr \left[ \hat{X} - X < -\frac{h}{2} \Big| X = t+h \right] dt \nonumber \\
&= \int_{0}^1 f_X(x) \Pr \left[ \hat{X} - x \ge \frac{h}{2} \Big| X = x \right] dx + \int_{-h}^{1-h} f_X(t+h) \Pr \left[ \hat{X} - t < \frac{h}{2} \Big| X = t+h \right] dt \nonumber \\
&\ge \int_{0}^{1-h} f_X(x) \Pr \left[ \hat{X} - x \ge \frac{h}{2} \Big| X = x \right] dx + \int_{0}^{1-h} f_X(t+h) \Pr \left[ \hat{X} - t < \frac{h}{2} \Big| X = t+h \right] dt \nonumber \\
&= \int_{0}^{1-h} f_X(x) \Pr \left[ \hat{X} - x \ge \frac{h}{2} \Big| X = x \right] + f_X(x+h) \Pr \left[ \hat{X} - x < \frac{h}{2} \Big| X = x+h \right] dx \nonumber \\
&= \int_{0}^{1-h} (f_X(x) + f_X(x + h)) \bigg\{ \frac{f_X(x)}{f_X(x) + f_X(x + h)} \Pr \left[ \hat{X} - x \ge \frac{h}{2} \Big| X = x \right] \nonumber \\
& ~~~ ~~~ ~~~ ~~~ ~~~ ~~~ ~~~ ~~~ ~~~ ~~~ ~~~ ~~~ ~~~ ~~~ ~~~ ~~~ + \frac{f_X(x+h)}{f_X(x) + f_X(x + h)} \Pr \left[ \hat{X} - x < \frac{h}{2} \Big| X = x + h \right] \bigg\} dx. \label{eq:long_eq2}
\end{align}
\end{figure*}

Notice that the quantity in the curly bracket implies the error probability of the decision rule 
\begin{align*}
\hat{X} - x \underset{X = x }{\overset{X=x+h}{\gtrless}} \frac{h}{2}.
\end{align*}
Then, it can be further bounded by the optimal error probability $P_{\textsf{min}}(x, x+h)$. Finally, we have the lower bound as follows. Let $f_{\textsf{ave}}(x, x+h):=f_X(x) + f_X(x + h)$. Then,
\begin{align*}
&\Pr \left[ |\hat{X}-X| \ge \frac{h}{2} \right] \ge \int_{0}^{1-h} (f_{\textsf{ave}}(x, x+h)) P_{\textsf{min}}(x, x+h) dx \\
&\implies \mathbb{E}[|X-\hat{X}|^r] \ge \\
&~~~ ~~~ \int_{0}^{\infty} r 2^{-r} h^{r-1} \int_{0}^{1-h} \frac{f_{\textsf{ave}}(x, x+h)}{2} P_{\textsf{min}}(x, x+h) dx dh \\
&= \int_{0}^{1} r 2^{-r} h^{r-1} \int_{0}^{1-h} \frac{f_{\textsf{ave}}(x, x+h)}{2} P_{\textsf{min}}(x, x+h) dx dh.
\end{align*}
It completes the proof.

\section{Inequality} \label{app:ineq}
The next proposition is a scalar version of $c_r$-inequality, which is originally for random variables, however, sufficient for our purpose.
\begin{proposition} \label{prop:power_r_UB}
For $a, b \ge 0$ and $r \in \mathbb{N}$,
\begin{align*}
(a+b)^r \le 2^r (a^r + b^r).
\end{align*}
\end{proposition}
\begin{IEEEproof}
By the binomial expansion theorem,
\begin{align*}
(a+b)^r &= \sum_{i=0}^r \binom{r}{i} a^i b^{r-i} \\
&\le \sum_{i=0}^r \binom{r}{i} \left(\max (a, b)\right)^r \\
&= 2^r \left(\max (a, b)\right)^r \\
&\le 2^r \left(a^r + b^r \right).
\end{align*}
\end{IEEEproof}

\end{document}